\documentclass[preprint2]{aastex}

\newcommand{\myemail}{siegfried.eggl@univie.ac.at}

\shorttitle{Habitable Zones in S-Type Binary Star Systems}
\shortauthors{Eggl et al.}

\begin{document}

\title{An Analytic Method to determine Habitable Zones for S-Type Planetary Orbits in Binary Star Systems}

\author{Siegfried Eggl\altaffilmark{1}, Elke Pilat-Lohinger\altaffilmark{1}, Nikolaos Georgakarakos\altaffilmark{2}, Markus Gyergyovits\altaffilmark{1} \and Barbara Funk\altaffilmark{1}}
\email{\myemail}

 \altaffiltext{1}{University of Vienna, Institute for Astronomy, T\"urkenschanzstr. 17, A-1180 Vienna, Austria}

 \altaffiltext{2}{128 V. Olgas str., Thessaloniki 546 45, Greece}

\begin{abstract}
With more and more extrasolar planets discovered in and around binary star systems, questions concerning the determination
of the classical Habitable Zone arise. Do the radiative and gravitational perturbations of the second star influence the extent of the Habitable Zone significantly, 
or is it sufficient to consider the host-star only? In this article we investigate the implications of stellar companions with different spectral types on the insolation a terrestrial planet
receives orbiting a Sun-like primary. We present time independent analytical estimates and compare these to insolation statistics gained via high precision numerical orbit calculations. 
Results suggest a strong dependence of permanent habitability on the binary's eccentricity, as well
as a possible extension of Habitable Zones towards the secondary in close binary systems.

\end{abstract}


\section{Introduction}

Fueled by the successes of 
current transit-observation incentives like KEPLER \citep{welsh-et-al-2012,borucki-koch-2011} and CoRoT \citep{corot-2009,corot-14b-2011} 
the quest for discovering the first Earth-twin 
has lead to a considerable cross-disciplinary interest
in the interplay between stellar and planetary properties to produce habitable worlds.
Even-though opinions differ on what exactly to look for in a system harboring a terrestrial planet in order to declare it 'habitable' 
(see e.g. \citet{buccino-et-al-2006}, \citet{kaltenegger-et-al-2007}, \citet{selsis-et-al-2007}, \citet{lammer-et-al-2009}), the classical
assumption investigated by \citet{kasting-et-al-1993}, i.e.      
the capacity for water to stay liquid on the planet's surface, may still be considered a prerequisite 
for the development and sustainability of complex life as we know it \citep{kaltenegger-sasselov-2011}.
This entails restrictions on planetological characteristics, such as mass, atmospheric and bulk composition, 
and sets limits to the host star's activity as well as radiation properties \citep{lammer-et-al-2009}.   
Dynamical considerations are of equal importance, since changes in orbital stability, 
or extreme variations in insolation due to large planetary eccentricities ($e_p > 0.7$) 
may also result in a hostile environment \citep{williams-pollard-2002}.
It is therefore only natural that one would look for a copy of our Solar System, when searching for habitable worlds.
Yet, the study of exoplanetary systems so far has clearly shown that a broader perspective is required.  

In fact, up to 70\% of all stellar systems in our galaxy may not be single but multi-stellar systems 
(e.g. \cite{kiseleva-eggleton-eggleton-2001} and references therein). Together with the 
approximately 60 planets that have already been discovered in systems harboring two stars \citep{schneider-2011} this suggests that
binary and multiple star systems should not be ignored in the search for habitable worlds.
 
Investigations of environments that permit planetary formation in binary star systems have progressed rapidly over the last decade (see e.g. \citet{thebault-2011} and references therein).  
Even-though important questions regarding the early phases of planet formation in binary star systems - especially the transition from planetesimal to planetary embryos - still remain to be answered,
late stage formation scenarios for terrestrial planets in such environments are available \citep{whitmire-et-al-1998,haghighipour-raymond-2007, haghighipour-et-al-2010}.
Since previous studies did consider the extent of the classical Habitable Zone (KHZ) to be purely a function of the primary star's luminosity and spectral type
as introduced by \citet{kasting-et-al-1993} (hereafter KWR93), we aim to refine this definition to encompass the 
gravitational and radiative influence of a second star.

This article is structured as follows: 
After a short recapitulation of the main radiative aspects of habitability as defined in KWR93 section \ref{sec:bp} introduces 
three exemplary binary-planet configurations, which will serve as test-cases for habitability considerations.  
Section \ref{sec:stab} briefly mentions the dynamical requirements which binary-planet configurations have to fulfill in order to ensure system stability.
In section \ref{sec:sec} the maximum radiative influence of the second star on a terrestrial planet in the primary's HZ is estimated and compared to actual 
insolation simulations. The occurring differences are being investigated in the following section.
Finally, generalized, analytical estimates are developed and compared to numerical simulations in sections \ref{sec:ana} \& \ref{sec:num},
and the results concerning the behavior of HZs in binary star systems are presented in section \ref{sec:res}. 
A discussion of the results concludes this article.

\section{Binary-Planet Configurations}
\label{sec:bp}
\begin{deluxetable}{ccc}
\tabletypesize{\scriptsize}
\tablecaption{Limiting radiation values for the inner ($A$) and outer ($B$) border of the HZ respectively in units of Solar constants ($1360\,[W/m^2]$). 
The values were taken from \citet{kasting-et-al-1993} assuming a runaway greenhouse scenario for the inner limit, and a maximum greenhouse effect for the outer limit.  \label{tab2}}
\tablewidth{0pt}
\tablehead{
\colhead{Spectral Type} & \colhead{$A$} & \colhead{$B$} 
}
\startdata
F0 & 1.90    & 0.46     \\
G2 & 1.41    & 0.36    \\
M0 & 1.05    & 0.27     
\enddata
\end{deluxetable} 

\begin{figure}
\plottwo{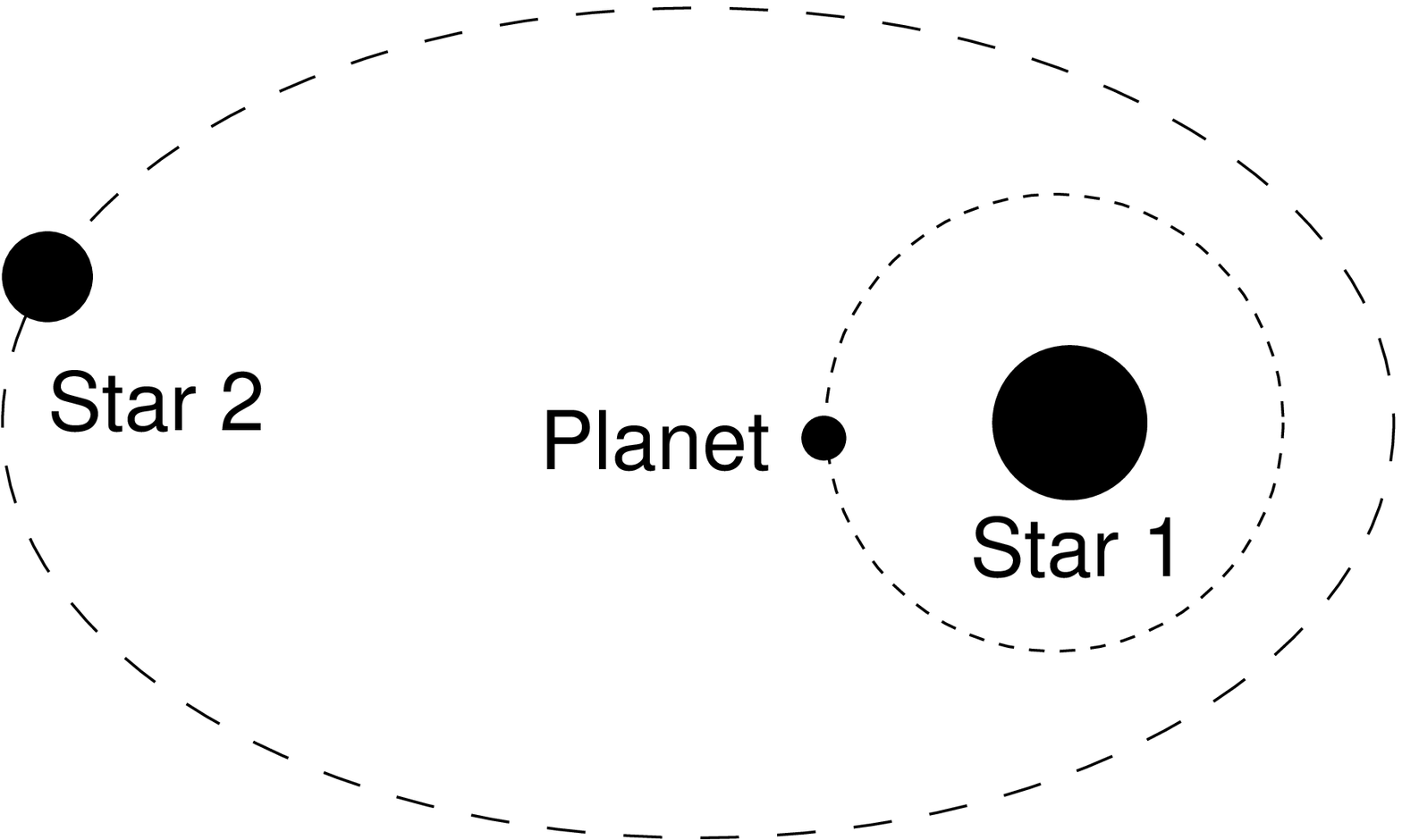}{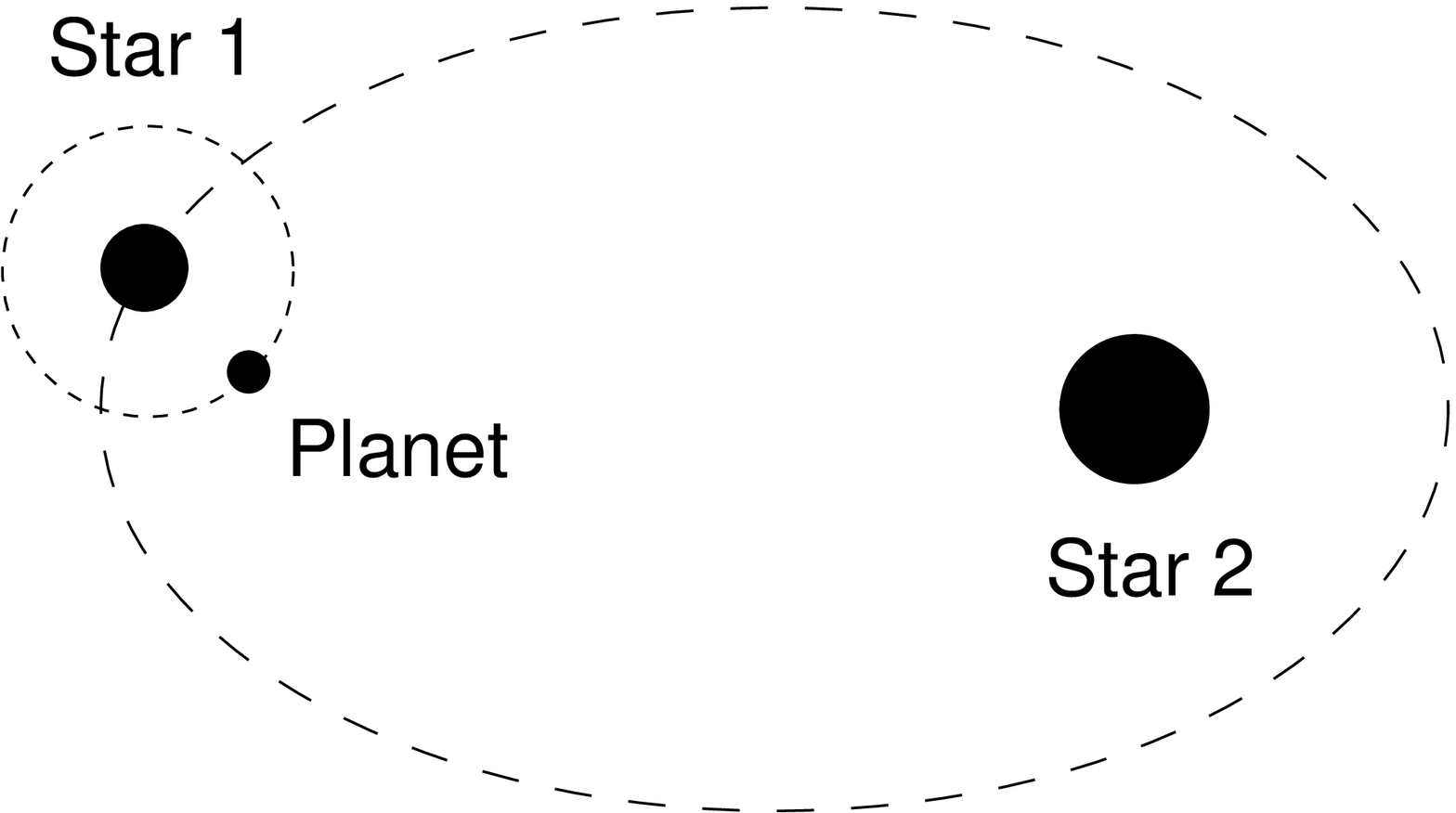}
\caption{Two examples of S-Type motion, i.e. the planet orbits only one binary-component \citep{dvorak-1984}; \textit{left}: S-Type I ($\mu \leq 0.5$), 
the more massive star is the planet's host (primary), 
\textit{right:} \mbox{S-Type II} ($\mu >0.5$), the less massive star is the planet's host.
 \label{fig1}}
\end{figure}

Apart from planetological and dynamical aspects, the insolation a terrestrial planet receives from its host star is naturally the main driver determining the extent of the HZ. 
When considering planets within a binary star system it is therefore important to track the combined radiation of both stars arriving at the planet.
KWR93 showed that not only the sheer amount of insolation, but also the spectral distribution is essential to estimate 
limiting values for atmospheric collapse. In order to model the impact of diverse stellar spectral classes on an Earth-like planet's atmosphere, 
KWR93 introduced so-called effective radiation values. These measure the relative impact a comparable 
amount of radiation (e.g. 1360 $[W/m^2]$) with different spectral properties has on a planet's atmosphere.   
Taking the effects of different stellar spectra into account is especially important in binary star systems with different stellar components.
Tab.~\ref{tab2} reproduces the effective radiation values for the inner (runaway greenhouse) and outer (maximum greenhouse) edge of a single star's HZ as given in KWR93.
Notice how the onset of runaway greenhouse effects requires almost twice as much radiation for a spectral distribution akin to F0 class stars compared to M0 spectral types.
Even though \citet{kaltenegger-sasselov-2011} assume similar effective radiation values for M and K spectral classes, actual calculations have only been done for F0, G2 and M0 ZAMS stars. 
For this reason, we first investigate the following three stellar configurations:
\begin{center}
\begin{tabular}{cccc}
i)& G2 - M0 &  $\mu \simeq 0.3$ \\
ii) & G2 - G2 &  $\mu \simeq 0.5$ \\
iii) & G2 - F0 & $\mu \simeq 0.6$
\end{tabular}
\end{center}
All stellar components are considered to be ZAMS stars, and $\mu=m_2/(m_1+m_2)$ denotes the binary's mass ratio.
A terrestrial planet is orbiting the Sun-like G2 host-star, hereafter referred to as primary.
Such binary-planet configurations are considered to be of satellite type (S-Type), i.e. the planet revolves around one star \citep{dvorak-1984}, see Fig.\,\ref{fig1}.
In fact, most of the planets in binary systems have been discovered to be of S-Type \citep{schneider-2011}, e.g. the system Gamma Cephei \citep{hatzes-et-al-2003}. 
In the following, we choose binary systems with semi-major axes between $10$ and $50\,AU$ for the comparison of numerical and analytical estimates on the extent of the HZ, 
as for closer binaries the HZs in G2-G2 and G2-F0 configurations are considerably reduced due to dynamical instability, whereas the qualitative differences for results beyond $50\,AU$ are small.
However, the methods presented in section \ref{sec:ana} are viable beyond those limits, as long as the assumptions given in \citet{georgakarakos-2003} remain valid.

The terrestrial planet is assumed to be cloudless \citep{kaltenegger-sasselov-2011} and orbits the G2 star only, 
resulting in configurations i) and ii) to be classified as S-Type I ($\mu \leq 0.5$) and iii) as S-Type II ($\mu >0.5$) respectively (see Fig.\,\ref{fig1}). 

\section{Dynamical Stability}
\label{sec:stab}
Binary-planet configurations that lead to highly chaotic orbits and eventual ejection of the planet cannot be considered habitable. Therefore dynamical stability of the system
is a basic requirement for our considerations.      
In order to assess the dynamical stability of a test planet's orbit within the given binary star systems, we applied results of numerical stability studies by \citet{rabl-dvorak-1988} (hereafter RD88), \citet{pilat-lohinger-dvorak-2002} (hereafter PLD02) and \citet{holman-wiegert-1999} 
(hereafter HW99) determining stable zones in a planar, normalized system (binary semi-major axis $a_{b}=1$).
In contrast to RD88 and HW99 who classified unstable orbits via ejections of test planets from the system, PLD02 applied the Fast Lyapunov Indicator (FLI) chaos detection method  \citep{froeschle-et-al-1997} implemented in a Bulirsch-Stoer extrapolation code
\citep{bulirsch-stoer-1964}.
In Tab.\,\ref{tab1} maximum planetary semi-major axes which still allow for a stable configuration in the normalized setup are shown for different binary orbits and mass ratios.
Results gained via FLI by PLD02 are compared to those published in HW99.
Even-though the critical values are very similar, the onset of dynamical chaos (PLD02) in G2-G2 and G2-F0 configurations appears for smaller planetary semi-major axes than predicted by the ejection criterion in HW99. 
For the following investigations the more conservative FLI stability estimates by PLD02 are used.
\begin{deluxetable}{ccccccc}
\tabletypesize{\scriptsize}
\tablecaption{Critical, planetary semi-major axis in a normalized binary star system $a_{b}=1$ with mass-ratio $\mu$ as stated in \citet{holman-wiegert-1999} compared 
to the values found by \citep{pilat-lohinger-dvorak-2002} via Fast Lyapunov chaos indicator (FLI). 
The higher the binary's eccentricity $e_b$, the smaller the test-planet's semi-major axis has to be to permit dynamical stability. 
Even-though the results are very similar, the FLI limits tend to be more conservative for higher mass ratios and binary eccentricities. \label{tab1}}
\tablewidth{0pt}
\tablehead{
& \multicolumn{2}{c}{$\mu=0.3$} & \multicolumn{2}{c}{$\mu=0.5$} & \multicolumn{2}{c}{$\mu=0.6$} \\
\colhead{$e_{b}$} & \colhead{FLI} & \colhead{HW99} & \colhead{FLI} & \colhead{HW99} & \colhead{FLI} & \colhead{HW99} 
}
\startdata
0.0 & 0.37 & 0.37 &  0.27 & 0.26 & 0.23   & 0.23 \\
0.1 & 0.29 & 0.30 &  0.25 & 0.24 & 0.21   & 0.20 \\
0.2 & 0.25 & 0.25 &  0.19 & 0.20 & 0.18 & 0.18 \\
0.3 & 0.21 & 0.21 &  0.16 & 0.18 & 0.15 & 0.16 \\ 
0.4 & 0.18 & 0.18 &  0.15 & 0.15 & 0.12 & 0.13 \\
0.5 & 0.13 & 0.14 &  0.12 & 0.12 & 0.09 & 0.10 \\
0.6 & 0.09 & 0.11 &  0.08 & 0.09 & 0.07 & 0.08 \\
0.7 & 0.07 & 0.07 &  0.05 & 0.06 & 0.05 & 0.05 \\
0.8 & 0.04 & 0.04 &  0.03 & 0.04 & 0.02 & 0.035 \\
0.9 & 0.01 &  -   &  0.01 & -    & 0.01 & -   
\enddata
\end{deluxetable}

\section{Influence of the Secondary}
\label{sec:sec}
The main question in dealing with habitability in binary star systems is doubtlessly: "How does the second star affect the HZ around the primary?"
Let us focus on the dynamical aspects first, as stability is a prerequisite to habitability.
The results from the previous section dictate that not all combinations of
a binary's semi-major axis $a_b$ and eccentricity $e_b$ are viable, if orbital stability of planets in the primary's HZ is required. 
Fig.~\ref{fig2}, \textit{top left} shows quadratic fits of the FLI stability data presented in Tab.~\ref{tab1}.
The secondary's allowed eccentricities are higher for planets orbiting the primary near the inner border of the KHZ, than 
for cases where the planet's semi-major axis is close to the KHZ's outer rim. Such dynamical restrictions set limits on how close
the secondary can approach the planet, which will in turn limit its insolation on the planet.
Applying analytical approximations introduced later in this section 
Fig.~\ref{fig2}, \textit{top right} demonstrates that the smallest possible separation between planet and secondary is always larger than the planet's aphelion distance to its primary.   
This lessens the second star's potential radiative contribution considerably for all but close S-Type II configurations.
In the latter case the enhanced luminosity of the secondary compensates larger distances to the planet. Using the minimum distances presented in Fig.~\ref{fig2}, \textit{top right}, one can estimate, 
that in a G2-G2 S-Type I system with $a_b=20AU$, the secondary's radiative influence
 on a planet started at the inner edge of the KHZ is in the order of only 10\% of the primary's contribution.
 As a consequence, the secondary was often considered not to have a significant radiative impact 
on the extent of the primary's HZ \citep{whitmire-et-al-1998, haghighipour-raymond-2007, haghighipour-et-al-2010}. 

Such a so-called "single-star approach", however, stands in contrast to numerical experiments presented in Fig.~\ref{fig3}, \textit{left}. The effective insolation curves shown were generated solving the full Newtonian 
Three-Body Problem (3BP) numerically via Lie-Series \citep{hanslmeier-dvorak-1984, eggl-dvorak-2010} and Gauss-Radau \citep{everhart-1974} integrators, where 
the actual amount of radiation the planet receives was calculated for each integration step.     
One can see that a planet started at the inner edge of the KHZ in a G2-G2 S-Type binary experiences an increase of more than 30\% in insolation compared to a planet on a circular orbit with corresponding semi-major axis around a single G2 star.    
Consequently, an Earth-like planet would spend a considerable time outside the classical, circular HZ in a G2-G2 configuration. 
Is, therefore, the secondary's radiative impact more important than assumed?
The fact that the variations in insolation perfectly correlate with the dynamical evolution of the planet's eccentricity (see Fig.~\ref{fig3}, \textit{right}) permits an alternative explanation: 
changes in the planetary orbit induced via gravitational perturbation by the second star might be responsible for the increased insolation values. Even-though planetary and binary orbits' semi-major axes are not expected to show any secular variation (e.g. \citet{harrington-1968}), 
it is known that even a distant companion would inject some eccentricity into the orbit of two bodies revolving around their common center of mass (e.g. \citet{mazeh-shaham-1979, georgakarakos-2002}). 
Elevated planetary eccentricities would entail smaller periastron distances - allowing for increased insolation by the primary. 

In order to draw a clearer picture on whether the perturbation induced rise in the planet's eccentricity or the secondary star's radiative influence are the dominant factors causing
the increased insolation onto the planet encountered in Fig.~\ref{fig3}, \textit{left}, we will make three assumptions which our analytical estimates will be based on: 
\begin{enumerate}
\item the binary-planet system is coplanar
%\item the planet moves on an orbit with time averaged eccentricity $\bar{e}_p=\left \langle e_p\right \rangle_t$
%\item wherever $\bar{e}_p$ is required, the variance of the planet's eccentricity is considered negligible so that 
%$\bar{e}_p = \sqrt{\langle e_p\rangle_t^2} \sim \sqrt{\langle e_p^2 \rangle_t}$. 
\item stellar luminosities $L$ are constant on timescales of the system's secular dynamics.
\item stellar occultation effects are negligible.
\end{enumerate}
Since we assume coplanar orbits, we will make extensive use of the analytic expressions in \citet{georgakarakos-2003, georgakarakos-2005}  to calculate the time averaged squared planetary eccentricity $\left \langle e^2_p\right \rangle_t$. 
In Appendix \ref{sec:emax} we will derive estimates on the planet's maximum injected eccentricity $e^{max}_p$. 
Unlike the recent ansatz by \citet{giuppone-et-al-2011} and earlier \citet{thebault-et-al-2006}, where the eccentricity evolution of a planet in a stellar binary was modeled by empirical formulae, 
\citet{georgakarakos-2003, georgakarakos-2005} derived an entirely analytical formula for the eccentricity of the inner - in this case the planet's - orbit of a hierarchical triple system, which is valid for a wide
range of mass ratios and initial conditions. Together with the estimates for $e^{max}_p$ in Appendix \ref{sec:emax}, the formulae presented in \citet{georgakarakos-2003, georgakarakos-2005} are an analytical
extension to the the first order with respect to the perturbing mass secular perturbation theory, as given for example in the longstanding work of \citet{heppenheimer-1978}.

Assumptions b) and c) are reasonable for well separated binary stars, where both components are on the main sequence.
The combination of the dynamical stability results presented in Fig.~\ref{fig2}, \textit{top left}, and the analytic expressions for the injected planetary eccentricity allow us to 
estimate not only the minimum distances between the secondary and the planet as shown in Fig.~\ref{fig2}, \textit{top right}, but also the maximum contributions to the planetary insolation from the primary and the secondary star respectively. 
%For this we require the planet's maximum eccentricity rather than its average.
% Identifying the approximate shape of the injected planetary eccentricities in Fig.~\ref{fig3}, \textit{right} as rectified sines, we will use the following approximation: 
% $\bar{e}_p \approx e^{max}_p/\sqrt{2}$.   

Fig.~\ref{fig2}, \textit{bottom left} shows the primary's effective insolation 
onto a terrestrial planet during its closest perihelion passage $d=a_p(1-e_p^{max})$. Here, 
the largest dynamically possible perturbation by the secondary is considered.
In order to separate the secondary's radiative and gravitational effects, the second star's radiation has been excluded in this plot.
Given a binary semi-major axis of $10AU$, Earth-analogues started at the inner edge of the KHZ permit considerably higher binary eccentricities 
(Fig.~\ref{fig2}, \textit{top left}) compared to planets near the outer edge.   
Therefore, higher insolations for planets near the inner edge of the KHZ are expected, 
since the injected planetary eccentricities are coupled strongly to the binary's eccentricity, see Appendix \ref{sec:emax}.
This effect abates around $a_b=20AU$,  where the primary's insolation becomes almost equal for planets started near the inner and outer rims of the KHZ.
For binary semi-major axes beyond $a_b=20AU$ planetary orbits closer to the secondary are more severely perturbed, than the ones near the inner edge of KHZ,
as the binaries' maximum allowed eccentricities become similar for inner and outer borders KHZ and the perturbation is merely dependent on the binary to planet period ratios.

 For planets on eccentric orbits in a G2-G2 configuration with $a_b=20AU$, the primary's insolation can increase up to almost 50\% compared to circular orbits sharing the same semi-major axis. In contrast, 
even with the planet being as close to the second star as dynamically possible, the secondary's maximum radiative input only accounts for an additional 10\% in the same setup. This can be seen in  Fig.~\ref{fig2}, \textit{bottom right}.
Hence, the primary is the main source of the additional insolation in S-Type I systems, which can only be explained via changes in the planet's eccentricity.   
Solely for S-Type II systems like G2-F0 binaries, the secondary's radiative contributions to planetary insolation can become comparable to the primary's (cf. Fig.~\ref{fig2}, \textit{bottom}).

\section{Consequences for the Habitable Zone}
\label{sec:def}    
In the previous section, the rise in the planet's eccentricity was identified as a main driver for the strong variance in simulated insolation curves in S-Type I systems.
Eccentric planetary orbits entail strong variations in insolation over an orbital period. Nevertheless, \citet{williams-pollard-2002} concluded, 
that this might not necessarily be prohibitive for habitability. As long as 
the average insolation values lie within habitable limits and $e_{p}<0.7$, the atmosphere should be able to act as a buffer, 
preventing immediate carbon freeze-out or instantaneous water evaporation.   
In order to distinguish between cases, where the planet remains within the radiative boundaries of the KHZ for all times, 
configurations where the planet is "on average" within the HZ, and non habitability, we introduce the following three categories:
\begin{description}
\item[Permanently Habitable Zone (PHZ): ]
The PHZ is the region where a planet always stays within the insolation limits ($A$, $B$) of the corresponding KHZ, i.e.  
\begin{equation}
A \geq S_{eff} \geq B
\end{equation}
where $A$ denotes the inner, and $B$ the outer effective radiation limit for a given spectral type, see Tab.~\ref{tab2}.

\item[Extended Habitable Zone (EHZ): ]
In contrast to the PHZ - where the planet stays within the KHZ for all times - parts of the planetary orbit lie outside the HZ 
due to e.g. the planet's eccentricity. Yet, the binary-planet configuration is still considered to be habitable when most of its orbit remains inside the HZ boundaries:
\begin{equation}
\left \langle S_{eff}\right \rangle_t +\sigma \leq A \quad \wedge \quad \left \langle S_{eff}\right \rangle_t -\sigma \geq B \label{eq:ehzlim}
\end{equation}
where  $\left \langle S_{eff} \right \rangle_t$ denotes the time-averaged effective insolation from both stars and $\sigma^2$ the effective insolation variance.   

\item[Averaged Habitable Zone (AHZ): ]
Following the argument of \citet{williams-pollard-2002}, this category encompasses all configurations 
which allow for the planet's time-averaged effective insolation to be within the limits of the KHZ.
\begin{equation}
 A \geq  \left \langle S_{eff} \right \rangle_t \geq B
\end{equation}

\end{description}
 
\section{Analytical Estimates}\label{sec:ana}    

We now propose analytical estimates to achieve a classification of planetary habitability as was suggested in the previous section. 
The aim is to circumvent time consuming numerical integrations
when global parameter scans are required to check systems for possible habitability.
Even-though the following analytical estimates are presented utilizing the $S_{eff}$ values developed in KWR93, more advanced 
atmospheric models for exoplanets can easily be introduced by exchanging the effective insolation values $A$ and $B$.
 
% Using the analytic expression in \citet{georgakarakos-2005}
% to identify the time averaged amplitude of the secular variations in the planet’s squared eccentricity $\left \langle e^2_p\right \rangle_t$,
% it is possible to derive analytical estimates for the PHZ, EHZ and AHZ under the following assumptions:
% 
% \begin{itemize}
% \item the planet moves on an orbit with constant eccentricity $\bar{e}_p=\left \langle e_p\right \rangle_t$
% \item wherever $\bar{e}_p$ is required, the variance of the planet's eccentricity is considered negligible so that 
% $\bar{e}_p = \sqrt{\langle e_p\rangle_t^2} \sim \sqrt{\langle e_p^2 \rangle_t}$. 
% \item stellar luminosities $L$ are constant on timescales of the system's secular dynamics.
% \end{itemize}
% The first two approximations are crude in the sense that the planet's actual orbit will of course change its eccentricity as a function of time, as can be
% seen in Fig.~\ref{fig3}. However, they will facilitate the computation 
% of radiation averages while still preserving relevant information on the dynamics of the system. 
% Fortunately for averages of squared eccentricity terms the second assumption can be dropped since $\langle e_p^2 \rangle_t$ 
% is given directly in \citet{georgakarakos-2005}. 
% The third assumption is reasonable for well separated binary stars, where both components are on the main sequence.
 
\subsection{Estimates for the PHZ}
\label{sec:PHZ}
Let the second star move on a fixed Keplerian orbit with semi-major axes $a_b$ and eccentricity $e_b$.
Accordingly, the planet's orbit has the semi-major axis $a_p$ and acquires a maximum eccentricity of $e_p^{max}$ due to the secondary's gravitational perturbations (cf. Fig~\ref{fig3}, \textit{right}).
This permits us to estimate the maximum and minimum insolation conditions for the planet to permanently remain within the KHZ:
\begin{description}
\item[insolation minimum condition:]
Both, planet and secondary are assumed to be in apocenter position and opposition. 
The additional normalization of the stellar luminosities per solid angle ($L=L_{bol}/(4\pi)$) 
with regard to the respective outer insolation limits for each star ($B_1$, $B_2$) ensures that different spectral properties are taken into account.
% Since all orbits are confined to the same plane, also occultation effects have to be considered.  
% \begin{displaymath}
% 1 \leq \left\{ \begin{array}{l} \Lambda > 0:\; \frac{L_1}{B_1} \left(a_p (1+\bar{e}_p)\right)^{-2}+ \\ 
%                 \qquad \quad\;\;                  \frac{\Lambda}{B_2} \left(a_b (1+e_b)+a_p(1+\bar{e}_p)\right)^{-2}  \\
% \Lambda \le 0:\; \frac{L_2}{ B_2} \left(a_p (1+\bar{e}_p)\right)^{-2} 
% \end{array} \right .
% \end{displaymath}
% where $\Lambda$ represents the secondary's remaining luminosity after accounting for the occultation by the primary star from the planet's perspective.   
% \begin{displaymath}
% \Lambda=\left(\varrho_2^2-\varrho_1^2\left(1+\frac{a_b(1+e_b)}{a_p(1+e_p)}\right)^2\right)\sigma T_2^4 
% \end{displaymath}
% Here, $\varrho$ denotes the respective stellar radius, $\sigma$ the Stefan-Boltzmann constant and $T$ the star's effective temperature. 
% If $\Lambda \le 0$, then the secondary is completely occulted by the primary. For the stellar configurations investigated in this work this is always the case.
% Therefore, it is sufficient to consider only the primary's insolation on the planet at apocenter.
\begin{displaymath}
1 \leq  \begin{array}{l} \frac{L_1}{B_1} \left(a_p (1+e^{max}_p)\right)^{-2}+ \\ 
                                \frac{L_2}{B_2} \left(a_b (1+e_b)+a_p(1+e^{max}_p)\right)^{-2} 
\end{array} 
\end{displaymath}
\item[insolation maximum condition:]
Again the luminosities are normalized, but this time with regard to the inner insolation limits ($A_1$, $A_2$).
Since we consider S-Type I and II systems, it is possible that the secondary at pericenter may produce a higher insolation on the planet than the primary star.
If this is the case, the maximum insolation configuration will have the planet at apocenter with regard to the primary, and the secondary at pericenter.
\begin{displaymath}
1  \geq  max \left\{ \begin{array}{l}
        \frac{L_1}{ A_1} \left(a_p (1-e^{max}_p)\right)^{-2} \\ +\frac{L_2}{ A_2} \left(a_b (1-e_b)-a_p (1-e^{max}_p)\right)^{-2}, \vspace{0.3 cm} \\
        \frac{L_1}{ A_1} \left(a_p(1+ e^{max}_p)\right)^{-2}\\ +\frac{L_2}{ A_2} \left(a_b (1-e_b)-a_p (1+e^{max}_p)\right)^{-2}
                      \end{array} \right .
\end{displaymath}
\end{description}

\subsection{Estimates for the AHZ}

The combined stellar insolation $S_{tot}$ on the planet is, of course, a function of time. 
In order to calculate time averaged insolation values, we will use that:
\begin{equation}
\left \langle S_{tot} \right \rangle_t = \left \langle S_{1} \right \rangle_t + \left \langle S_{2} \right \rangle_t 
\end{equation}
where $\left \langle S_1 \right \rangle_t$ is the time average of the planetary insolation due to its host star,
and $\left \langle S_{2} \right \rangle_t$ the time averaged contribution of the second star.
Let us focus on the two-body problem planet - host star first:
Here, $\left \langle S_1 \right \rangle_t = \left \langle L_1/\delta^2(t) \right \rangle_t$ 
where $\delta(t)$ denotes the scalar distance between planet and primary. 
The average insolation a planet on an unperturbed Keplerian orbit experiences can be calculated using the planet's angular momentum $h=\delta^2\dot{f}$,
 $f$ being the true anomaly.
\begin{eqnarray}
\langle S_1 \rangle_t &=&\frac{L_1}{P}\int_0^P \frac{1}{\delta^2(t)}dt \\ \nonumber
&=&\frac{L_1 n}{2\pi} \int_0^{2\pi} \frac{1}{h}df \\ \nonumber
        &=& \frac{L_1 n}{ h} 
\end{eqnarray}
This expression states that the time-averaged insolation a planet receives over one orbit depends only on the star's luminosity, the planet's mean motion $n$ 
and its orbital angular momentum (see e.g. \citet{seager-2010}, p.~18). 
Now let us construct a circular orbit so that $h_{circle} = h_{ellipse}$.
A planet moving on such an orbit with 'equivalence radius' $r=a_p(1-\langle e_p^2 \rangle_t)$ will 
experience the same amount of insolation per unit time as a planet on an elliptic orbit sharing the same angular momentum. 
The advantage of considering  "equivalent circular orbits" is that the insolation remains constant for any orbital position of the planet.
Of course, the reduction of elliptic orbits to circular orbits with common angular momentum will decrease the planet's orbital period, 
as $r\leq a$ and $r=a$ only if $\langle e_p^2 \rangle_t = 0$. 
However, as we have chosen the 'equivalent circular orbit' to have the same angular momentum, it will share the same constant rate of change of insolation with the true orbit.
Therefore, even an average over the longer period of the elliptic orbit will still yield valid results.  

%the insolation on a circular orbit is independent of time however\footnote{given the assumption that the stellar luminosities are constant on orbital timescales}, the insolation average of the circular orbit over the circular orbital period equals the average over the slightly longer 
%elliptic period. Therefore the change in period following from the equivalent circularization of the orbit is of no consequence for our purposes.

\begin{figure}[h]
\epsscale{0.5}
\plotone{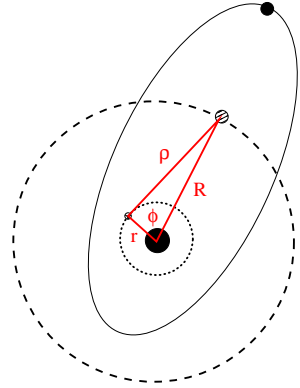}
\caption{'Equivalence Orbits' of the planet (dotted) and the secondary (dashed) around the primary. 
$R$ denotes the equivalence radius of the secondary's circular orbit sharing the same angular momentum with its actual elliptic orbit (continuous).
The planet's equivalence radius ($r$), the distance between the time averaged secondary's and planet's positions ($\rho$)
as well as the angle ($\phi$) opposite to ($\rho$) are highlighted.
In order to calculate the secondary's mean insolation on the planet, an averaging over $1/\rho^2(\phi)$ is required.}\label{fig4}
\end{figure}

We will now apply the same line of argumentation to construct the equivalence radius $R$ for the secondary. 
This allows us to extend the pseudo-static radiation environment 
to include the average radiative influence of the secondary. 
Since the distinction between averaged and initial eccentricity is small for the secondary - its 
secular variance is negligible for the cases investigated - we can safely use the secondary's initial eccentricity to calculate $R$.
The suggested configuration is shown in Fig.~\ref{fig4}. The secondary circles the primary in a fictitious orbit with equivalence radius $R$, and the planet orbits the primary 
in a circle with radius $r$.

The insolation on the planet depends on the relative distances of the planet to both stars.
In order to estimate the insolation on the planet caused by the secondary ($S_2$), we simply apply law of cosine, and average over all possible geometric configurations (Fig.~\ref{fig4}):
\begin{eqnarray}
\rho^2 &= &R^2+r^2-2 r R \cos(\phi) \nonumber \\ 
\left \langle S_2 \right \rangle_t &= & \left \langle \frac{L_2}{ \rho^2(\phi)} \right \rangle_\phi \\
 &= &  \frac{L_2}{2\pi}\int_0^{2\pi}\frac{1}{R^2+r^2-2 r R \,\cos(\phi)}\,d\phi \nonumber \\
  &= &\frac{L_2}{(R^2-r^2)}\quad if \; R>r \nonumber 
  \end{eqnarray}
where $\rho$ is the planet's distance to the secondary, and $\phi$ denotes the angle between the distance vectors of planet and secondary to the host star.
Since we do not allow for orbit crossings of the planet and the secondary, $R>r$ will always hold.
Consequently, the total, time averaged insolation onto the planet is given by:
 \begin{equation} 
\left \langle S_{tot} \right \rangle_t = \frac{L_1}{r^2}+\frac{L_2}{(R^2-r^2)} \label{eq:av}
  \end{equation}
Equation (\ref{eq:av}) does not yet take the different spectral properties of the binary's components into account. 
Therefore we note the following conditions for the planet's averaged insolation being within habitable limits:

\begin{description}
\item[average insolation minimum condition:]
\begin{displaymath}
1 \leq  \left \langle S_{eff, B} \right \rangle_t =  \frac{L_1}{ B_1}\frac{1}{ r^2}+\frac{L_2}{ B_2}\frac{1}{R^2-r^2}  
\end{displaymath}
\item[average insolation maximum condition:]
\begin{displaymath}
1 \geq \left \langle S_{eff, A} \right \rangle_t =  \frac{L_1}{ A_1}\frac{1}{r^2}+\frac{L_2}{ A_2}\frac{1}{R^2-r^2} 
\end{displaymath}
\end{description}
Here, the indices $1,2$ indicate the respective star's KHZ boundary values $A$ and $B$, 
$R=a_b(1-e_b^2)$ and $r=a_p(1-\langle e_p^2 \rangle_t)$ where the averaged squared planetary eccentricity was calculated following
\citet{georgakarakos-2005}. 
 
\subsection{Estimates for the EHZ}
\label{subsec:EHZ}
Having derived the insolation time averages in the previous section, the expected insolation variance ($\sigma^2$) still remains to be determined. 
\begin{equation}
\sigma^2=\langle S_{tot}^2\rangle_t- \langle S_{tot}\rangle_t^2
\end{equation}
Using equation (\ref{eq:av}), and analytic estimates for $\langle S_{tot}^2 \rangle_t$, 
the effective insolation variance can be calculated as follows (see Appendix \ref{sec:var}):
\begin{eqnarray}
\sigma^2_{X} &=&\frac{L_1^2}{X_1^2r^4}\left(-1+3\langle e_p^2\rangle-3\langle e_p^2\rangle^2+\langle e_p^2\rangle^3\right) \nonumber\\ 
            &&+\frac{L_1^2}{X_1^2r^4}\sqrt{1-\langle e_p^2\rangle}\left(1-\frac{\langle e_p^2\rangle}{2} - \frac{\langle e_p^2\rangle^2}{2} \right) \nonumber \\ 
            &&-\frac{2L_1L_2}{X_1X_2(r^4-r^2R^2)}\left(1-\sqrt{1-\langle e_p^2\rangle}\left(1+\langle e_p^2\rangle\right)\right) \nonumber \\
            &&-\frac{2L_2^2r^2}{X_2^2(r^2-R^2)^3}
\end{eqnarray}
 where $X_i \in \{A_i,B_i\}$ and the index $i$ denotes the respective star. The minimum and maximum conditions for a planet to be within the EHZ are given as:
\begin{description}
\item[extended insolation minimum condition:]
\begin{displaymath}
1 \leq  \left \langle S_{eff, B} \right \rangle_t -\sigma_B  
\end{displaymath}
\item[extended insolation maximum condition:]
\begin{displaymath}
1 \geq \left \langle S_{eff, A} \right \rangle_t +  \sigma_A 
\end{displaymath}
\end{description}

\section{Reliability of Analytical Estimates}
\label{sec:num}

\begin{deluxetable}{c|cccc|cccc|cccc|}
\tabletypesize{\scriptsize}
%\rotate
\tablecaption{Percentages of planetary orbits classified identically via a numerical simulations and analytical estimates as presented in section \ref{sec:ana}. Three binary component configurations have been investigated, 
the reference classifications were extracted from numerical orbit integrations and insolation simulations. 
\label{tab3}}
\tablewidth{0pt}
\tablehead{
$[AU]$& \multicolumn{4}{c}{G2-M0 $[\%]$} & \multicolumn{4}{c}{G2-G2 $[\%]$} & \multicolumn{4}{c}{G2-F0 $[\%]$}\\
\colhead{$a_{b}$} & \colhead{\textbf{Total}} & \colhead{PHZ} & \colhead{EHZ} & \colhead{AHZ}& \colhead{\textbf{Total}} & \colhead{PHZ} & \colhead{EHZ} & \colhead{AHZ} & \colhead{\textbf{Total}} & \colhead{PHZ}& \colhead{EHZ} & \colhead{AHZ}
}
\startdata
10 & \textbf{ 95.9} &  97.4  &  99.8  &  98.3  & \textbf{ 94.4} &  97.4  &  99.2  &  98.0  & \textbf{ 93.6} &  95.8  &  98.5  &  98.2  \\  
20 & \textbf{ 98.8} &  99.3  &  99.5  &  99.8  & \textbf{ 98.5} &  99.2  &  99.6  &  99.6  & \textbf{ 98.5} &  99.5  &  99.4  &  99.6  \\  
30 & \textbf{ 99.0} &  99.5  &  99.7  &  99.9  & \textbf{ 99.2} &  99.7  &  99.6  &  99.8  & \textbf{ 98.9} &  99.8  &  99.4  & 100.0  \\  
40 &\textbf{ 99.2} &  99.5  &  99.6  &  99.9  & \textbf{ 99.3} &  99.9  &  99.6  & 100.0  & \textbf{ 99.0} &  99.8  &  99.5  &  99.8   \\  
50 &\textbf{ 99.2} &  99.6  &  99.7  &  99.9  & \textbf{ 99.4} &  99.7  &  99.7  &  99.9  & \textbf{ 99.4} &  99.8  &  99.7  &  99.9   \\  
\enddata
\end{deluxetable}

In order to test the reliability of the analytical estimates presented in section~\ref{sec:ana}, we used high precision 
numerical integration methods based on Gauss Radau quadrature \citep{everhart-1974} and Lie Series
\citep{hanslmeier-dvorak-1984, eggl-dvorak-2010} to determine the actual positions of both stars with respect to the 
Earth-like planet. Assuming that the stellar luminosities will not change significantly 
on the timescale of the planet's secular period in eccentricity, 
good approximations of insolation patterns can be obtained in such a way. 
Fig.~\ref{fig5} shows a comparison of analytic habitability classifications 
versus results gained via numerical orbit integration and direct insolation calculation.
The setup consists of a terrestrial planet ($1\,M_\oplus$) in S-Type orbit around a G2 host star, with three different spectral types as secondary: 
F0 (\textit{top}), G2 (\textit{mid}) and M0 (\textit{bottom}).
The terrestrial planet was started on circular orbits with semi-major axes between $0.6\,AU \le a_{p} \le 2\,AU$ with the secondaries' semi-major axes being $a_{b}=50\,AU$. 
The time span of the numerical integrations encompassed at least two secular periods in the planet's eccentricity.  
It is evident that for small binary eccentricities, all three types of HZ coincide well with the borders defined in KWR93 indicated by the vertical lines at 
$0.84$ and $1.67\,AU$. For $e_{b}>0.1$ however, a splitting into the HZ categories defined in sections~\ref{sec:def} and \ref{sec:ana} becomes eminent. The PHZ (black - blue online)
shrinks considerably with growing eccentricity of the binary's orbit. This is due to the perturbation induced elevation of the planet's eccentricity.
In contrast, the region defined in KWR93 is best approximated by the AHZ (light grey - yellow online), which remains virtually unaffected by the secondary's eccentricity.
In this setup all analytically calculated HZs are in excellent agreement with the numerical ones.
Only close to the stability limit (shaded region - purple in the online version) the correspondence between simulation and analytical estimates decreases.
This can be seen more clearly when the secondary's influence becomes stronger, e.g. in the cases of $a_{b}=10\,AU$, see Fig.~\ref{fig6}. 
% The last row in Fig. \ref{fig5} presents the classification differences of analytical and numerical results. One can see that the fundamental behavior is captured well,
% as about $91\%$ of the orbits for $a_{secondary}=10\,AU$ and $98\%$ for $a_{secondary}=30\,AU$ were classified correctly. 
In general the analytical approach is producing more conservative results compared to the numerical data (cf. Figs.~\ref{fig7}~\&~\ref{fig6}).
In order to determine whether these deviations in the determination of the PHZ are due to
\begin{enumerate}
\item inaccurate analytical estimates for $e_p^{max}$,
\item insufficient time resolution in numerical simulations with regard to determining $e_p^{max}$ values, or
\item insufficient total integration time to reach minimum and maximum insolation conditions, 
\end{enumerate}
we constructed semi-analytical PHZs using numerically determined $e_p^{max}$ values in the analytic equations to determine the PHZ presented in section \ref{sec:PHZ}.   
The borders of the semi-analytically derived PHZs are depicted as white dashed-dotted lines in Figs.~\ref{fig7}~\&~\ref{fig6}.
As they are nearly identical with the fully analytic estimates, we can exclude a) and b), which would have lead to significantly different results 
for semi-analytic and analytic approaches. 
Therefore, c) seems most likely to cause the differences between the numerical and analytical PHZs, since encountering an exact alignment of planetary aphelion and secondary perihelion at the moment where $e_p=e_p^{max}$ may take far longer than two secular periods.  
As the computational efforts required to ensure that a simulation's time resolution as well as total integration time are sufficient to identify the correct PHZ boundaries 
are enormous, the necessity to have analytical methods at hand becomes evident.
 
As far as EHZ and AHZ regions are concerned, clear differences can be seen in high perturbation environments close to the transition to instability (Figs.~\ref{fig7}~\&~\ref{fig6}, shaded regions). 
In these cases, the authors favor the numerical results, as single configurations are not critical for the more statistically oriented measures.

% deviations of the total simulation time from exact multiples of secular periods may produce difference results for the numerically derived AHZ and EHZ, 
  
A quantitative overview of the correspondence between numerical and analytical results for all system configurations investigated is given in Tab.~\ref{tab3}. 
Here, similar maps as presented in Figs.~\ref{fig5}~\&~\ref{fig6} were generated with a resolution of $\Delta a_p=0.01\,AU$,
and $\Delta e_b=0.01$ and evaluated statistically. 
In spite of their shortcomings in determining the PHZ, the numerical classification results have been used as reference values and are compared to the analytical estimates given in section \ref{sec:ana}. 
The total correspondence percentages are  
calculated as the number of all orbits below the stability limit that were classified identically via numerics and analyitcs, divided by the total number of 
orbits simulated. The number of orbits classified as PHZ analytically divided by the number of orbits classified as PHZ numerically yields
the percentage of PHZ, etc. Tab.~\ref{tab3} shows that the global correspondence between both methods is quite convincing, 
which can be considered a strong indicator that the behavior of the respective HZs is modeled correctly.  

Also, Figs.~\ref{fig7},~\ref{fig5}~\&~\ref{fig6} indicate that most of the significant deviations between numerical and analytical results occur near the border of orbital instability,  
especially for high mass and small period ratios. 
This might be expected for AHZs and EHZs given the approximations involved in determining the analytic estimates. 
In the case of PHZs, however, semi-analytical results suggest caution in using simulation outcomes as reference values.

\section{Results}
\label{sec:res}
Diverse trends in the behavior of the different types of HZs can be seen in Figs.~\ref{fig5}~\&~\ref{fig6}.  
While the AHZ is almost independent of the binary's eccentricity and coincides well with the KHZ by KWR93 for distant stellar companions, 
the PHZ and EHZ shrink with higher binary eccentricities.
The fact that the PHZ and EHZ contract around the center of the KHZ emphasizes the importance of the changes in the 
planet's eccentricity, as the secondary's radiation alone could only account for one-sided features towards the outer edge of the KHZ.     
The PHZs seem most affected by strong perturbations and shrink to almost half their size before the systems investigated become unstable.  
Interestingly, in close S-Type II binaries with low eccentricity the extent of the EHZs and AHZs can reach beyond the predicted values by KWR93.
This can be seen in Fig.~\ref{fig7}, where a zoom on the outer border of the KHZ in a G2-F0 configuration is shown. 
Here an extension towards the second star of about $0.1\,AU$ seems possible for the system's AHZ if the binary's eccentricity 
remains below the system's stability limit of $e_b \simeq 0.2$.

%Figure 5,6

\section{Conclusions}
In this work, the impact of the second star on Habitable Zones in S-Type binary star systems with different stellar constituents 
has been investigated analytically as well as numerically.
The radiative contribution of the secondary on a terrestrial planet is negligible in all but S-Type II systems, 
if orbital stability of the planet is required.
The gravitational influence of the second star on the other hand perturbs in the planet's eccentricity, 
which in turn can lead to substantial changes in planetary insolation. 
Therefore the secondary has indeed to be taken into account when calculating the extent of Habitable Zones.  
 
Our analytical estimates for planetary eccentricities in binary star systems introduced in Appendix \ref{sec:emax} are an extension 
to secular perturbation theory as used in e.g. \citet{heppenheimer-1978}.
Together with methods presented in section \ref{sec:ana} suggesting an analytic determination of Habitable Zones, 
they allow to paint a global picture of habitability in \mbox{S-Type} binary star systems without having to rely on time consuming numerical orbit integrations.
Our approach is quite flexible in the sense that different planetary atmospheric models and average stellar 
luminosities can be integrated via adaption of the $S_{eff}$ values. 
Thus, the formulae presented in this article grant access to calculating Habitable Zones for a large set of possible binary-planet configurations.

For the three stellar configurations investigated it could be shown that the Permanently Habitable Zone, i.e. the zone where the planet never exceeds the classical insolation limits for habitability shrinks considerably with the binary's eccentricity. 
If one considers average insolation values only, the extent of the Average Habitable Zone coincides well with predictions by \citet{kasting-et-al-1993} for wide binaries, whereas a
significant extension towards the secondary is possible for close, eccentric binary systems.
The overall correspondence between numerical and analytical results presented is excellent, as 93-99\% of all investigated orbits were classified identically. 
The computational efforts required to calculate the true extent of Permanently Habitable Zones numerically, however, can be enormous and might in fact be prohibitive in some cases.
In contrast, the analytical method presented offers immediate, reliable estimates.   

A more careful approach than the one proposed in this work is advisable when multiplanetary systems, systems close to the stability limit or resonant configurations are being investigated.
In a next step we plan to extend our classification methods to mutually inclined systems.

\acknowledgments

The authors would like to acknowledge the support of FWF projects AS11608-N16 (EP-L, SE), P20216-N16 (SE, MG \& EP-L) and P22603-N16 (EP-L \& BF). SE acknowledges the support of 
University of Vienna's Forschungsstipendium 2012.

\bibliographystyle{apj}
%\bibliography{/home/eggl/Uni/biblio/eggl}

\begin{thebibliography}{36}
\expandafter\ifx\csname natexlab\endcsname\relax\def\natexlab#1{#1}\fi

\bibitem[{{Baglin} {et~al.}(2009){Baglin}, {Auvergne}, {Barge}, {Deleuil},
  {Michel}, \& {The CoRoT Exoplanet Science Team}}]{corot-2009}
{Baglin}, A., {Auvergne}, M., {Barge}, P., {Deleuil}, M., {Michel}, E., \& {The
  CoRoT Exoplanet Science Team}. 2009, in IAU Symposium, Vol. 253, IAU
  Symposium, 71--81

\bibitem[{{Borucki} \& {Koch}(2011)}]{borucki-koch-2011}
{Borucki}, W.~J., \& {Koch}, D.~G. 2011, in IAU Symposium, Vol. 276, IAU
  Symposium, ed. {A.~Sozzetti, M.~G.~Lattanzi, \& A.~P.~Boss}, 34--43

\bibitem[{{Buccino} {et~al.}(2006){Buccino}, {Lemarchand}, \&
  {Mauas}}]{buccino-et-al-2006}
{Buccino}, A.~P., {Lemarchand}, G.~A., \& {Mauas}, P.~J.~D. 2006, Icarus, 183,
  491

\bibitem[{Bulirsch \& Stoer(1964)}]{bulirsch-stoer-1964}
Bulirsch, R., \& Stoer, J. 1964, Numerische Mathematik, 6, 413,
  10.1007/BF01386092

\bibitem[{{Dvorak}(1984)}]{dvorak-1984}
{Dvorak}, R. 1984, Celestial Mechanics, 34, 369

\bibitem[{{Eggl} \& {Dvorak}(2010)}]{eggl-dvorak-2010}
{Eggl}, S., \& {Dvorak}, R. 2010, in Lecture Notes in Physics, Berlin Springer
  Verlag, Vol. 790, Lecture Notes in Physics, Berlin Springer Verlag, ed.
  {J.~Souchay \& R.~Dvorak}, 431--480

\bibitem[{{Everhart}(1974)}]{everhart-1974}
{Everhart}, E. 1974, Celestial Mechanics, 10, 35

\bibitem[{{Froeschl{\'e}} {et~al.}(1997){Froeschl{\'e}}, {Lega}, \&
  {Gonczi}}]{froeschle-et-al-1997}
{Froeschl{\'e}}, C., {Lega}, E., \& {Gonczi}, R. 1997, Celestial Mechanics and
  Dynamical Astronomy, 67, 41

\bibitem[{{Georgakarakos}(2002)}]{georgakarakos-2002}
{Georgakarakos}, N. 2002, \mnras, 337, 559

\bibitem[{{Georgakarakos}(2003)}]{georgakarakos-2003}
---. 2003, \mnras, 345, 340

\bibitem[{{Georgakarakos}(2005)}]{georgakarakos-2005}
---. 2005, \mnras, 362, 748

\bibitem[{{Giuppone} {et~al.}(2011){Giuppone}, {Leiva}, {Correa-Otto}, \&
  {Beaug{\'e}}}]{giuppone-et-al-2011}
{Giuppone}, C.~A., {Leiva}, A.~M., {Correa-Otto}, J., \& {Beaug{\'e}}, C. 2011,
  \aap, 530, A103

\bibitem[{{Haghighipour} {et~al.}(2010){Haghighipour}, {Dvorak}, \&
  {Pilat-Lohinger}}]{haghighipour-et-al-2010}
{Haghighipour}, N., {Dvorak}, R., \& {Pilat-Lohinger}, E. 2010, in Astrophysics
  and Space Science Library, Vol. 366, Astrophysics and Space Science Library,
  ed. {N.~Haghighipour}, 285--+

\bibitem[{{Haghighipour} \& {Raymond}(2007)}]{haghighipour-raymond-2007}
{Haghighipour}, N., \& {Raymond}, S.~N. 2007, \apj, 666, 436

\bibitem[{{Hanslmeier} \& {Dvorak}(1984)}]{hanslmeier-dvorak-1984}
{Hanslmeier}, A., \& {Dvorak}, R. 1984, Astronomy and Astrophysics, 132, 203

\bibitem[{{Harrington}(1968)}]{harrington-1968}
{Harrington}, R.~S. 1968, \aj, 73, 190

\bibitem[{{Hatzes} {et~al.}(2003){Hatzes}, {Cochran}, {Endl}, {McArthur},
  {Paulson}, {Walker}, {Campbell}, \& {Yang}}]{hatzes-et-al-2003}
{Hatzes}, A.~P., {Cochran}, W.~D., {Endl}, M., {McArthur}, B., {Paulson},
  D.~B., {Walker}, G.~A.~H., {Campbell}, B., \& {Yang}, S. 2003, Astrophysical
  Journal, 599, 1383

\bibitem[{{Heppenheimer}(1978)}]{heppenheimer-1978}
{Heppenheimer}, T.~A. 1978, \aap, 65, 421

\bibitem[{{Holman} \& {Wiegert}(1999)}]{holman-wiegert-1999}
{Holman}, M.~J., \& {Wiegert}, P.~A. 1999, \aj, 117, 621

\bibitem[{{Kaltenegger} \& {Sasselov}(2011)}]{kaltenegger-sasselov-2011}
{Kaltenegger}, L., \& {Sasselov}, D. 2011, \apjl, 736, L25

\bibitem[{{Kaltenegger} {et~al.}(2007){Kaltenegger}, {Traub}, \&
  {Jucks}}]{kaltenegger-et-al-2007}
{Kaltenegger}, L., {Traub}, W.~A., \& {Jucks}, K.~W. 2007, \apj, 658, 598

\bibitem[{{Kasting} {et~al.}(1993){Kasting}, {Whitmire}, \&
  {Reynolds}}]{kasting-et-al-1993}
{Kasting}, J.~F., {Whitmire}, D.~P., \& {Reynolds}, R.~T. 1993, Icarus, 101,
  108

\bibitem[{{Kiseleva-Eggleton} \&
  {Eggleton}(2001)}]{kiseleva-eggleton-eggleton-2001}
{Kiseleva-Eggleton}, L., \& {Eggleton}, P.~P. 2001, in Astronomical Society of
  the Pacific Conference Series, Vol. 229, Evolution of Binary and Multiple
  Star Systems, ed. {P.~Podsiadlowski, S.~Rappaport, A.~R.~King, F.~D'Antona,
  \& L.~Burderi }, 91

\bibitem[{{Lammer} {et~al.}(2009){Lammer}, {Bredeh{\"o}ft}, {Coustenis},
  {Khodachenko}, {Kaltenegger}, {Grasset}, {Prieur}, {Raulin}, {Ehrenfreund},
  {Yamauchi}, {Wahlund}, {Grie{\ss}meier}, {Stangl}, {Cockell}, {Kulikov},
  {Grenfell}, \& {Rauer}}]{lammer-et-al-2009}
{Lammer}, H., {et~al.} 2009, \aapr, 17, 181

\bibitem[{{Mazeh} \& {Shaham}(1979)}]{mazeh-shaham-1979}
{Mazeh}, T., \& {Shaham}, J. 1979, \aap, 77, 145

\bibitem[{{Pilat-Lohinger} \& {Dvorak}(2002)}]{pilat-lohinger-dvorak-2002}
{Pilat-Lohinger}, E., \& {Dvorak}, R. 2002, Celestial Mechanics and Dynamical
  Astronomy, 82, 143

\bibitem[{{Rabl} \& {Dvorak}(1988)}]{rabl-dvorak-1988}
{Rabl}, G., \& {Dvorak}, R. 1988, \aap, 191, 385

\bibitem[{Schneider(2011)}]{schneider-2011}
Schneider, J. 2011, The Extrasolar Planets Encyclopaedia

\bibitem[{{Seager}(2010)}]{seager-2010}
{Seager}, S. 2010, {Exoplanets}, ed. {Seager, S.}

\bibitem[{{Selsis} {et~al.}(2007){Selsis}, {Kasting}, {Levrard}, {Paillet},
  {Ribas}, \& {Delfosse}}]{selsis-et-al-2007}
{Selsis}, F., {Kasting}, J.~F., {Levrard}, B., {Paillet}, J., {Ribas}, I., \&
  {Delfosse}, X. 2007, \aap, 476, 1373

\bibitem[{{Thebault}(2011)}]{thebault-2011}
{Thebault}, P. 2011, Celestial Mechanics and Dynamical Astronomy, 111, 29

\bibitem[{{Th{\'e}bault} {et~al.}(2006){Th{\'e}bault}, {Marzari}, \&
  {Scholl}}]{thebault-et-al-2006}
{Th{\'e}bault}, P., {Marzari}, F., \& {Scholl}, H. 2006, \icarus, 183, 193

\bibitem[{{Tingley} {et~al.}(2011){Tingley}, {Endl}, {Gazzano}, {Alonso},
  {Mazeh}, {Jorda}, {Aigrain}, {Almenara}, {Auvergne}, {Baglin}, {Barge},
  {Bonomo}, {Bord{\'e}}, {Bouchy}, {Bruntt}, {Cabrera}, {Carpano}, {Carone},
  {Cochran}, {Csizmadia}, {Deleuil}, {Deeg}, {Dvorak}, {Erikson},
  {Ferraz-Mello}, {Fridlund}, {Gandolfi}, {Gillon}, {Guenther}, {Guillot},
  {Hatzes}, {H{\'e}brard}, {L{\'e}ger}, {Llebaria}, {Lammer}, {Lovis},
  {MacQueen}, {Moutou}, {Ollivier}, {Ofir}, {P{\"a}tzold}, {Pepe}, {Queloz},
  {Rauer}, {Rouan}, {Samuel}, {Schneider}, {Shporer}, \&
  {Wuchterl}}]{corot-14b-2011}
{Tingley}, B., {et~al.} 2011, \aap, 528, A97

\bibitem[{{Welsh} {et~al.}(2012){Welsh}, {Orosz}, {Carter}, {Fabrycky}, {Ford},
  {Lissauer}, {Prsa}, {Quinn}, {Ragozzine}, {Short}, {Torres}, {Winn}, {Doyle},
  {Barclay}, {Batalha}, {Bloemen}, {Brugamyer}, {Buchhave}, {Caldwell},
  {Caldwell}, {Christiansen}, {Ciardi}, {Cochran}, {Endl}, {Fortney},
  {Gautier}~III, {Gilliland}, {Haas}, {Hall}, {Holman}, {Howard}, {Howell},
  {Isaacson}, {Jenkins}, {Klaus}, {Latham}, {Li}, {Marcy}, {Mazeh}, {Quintana},
  {Robertson}, {Shporer}, {Steffen}, {Windmiller}, {Koch}, \&
  {Borucki}}]{welsh-et-al-2012}
{Welsh}, W.~F., {et~al.} 2012, Nature

\bibitem[{{Whitmire} {et~al.}(1998){Whitmire}, {Matese}, {Criswell}, \&
  {Mikkola}}]{whitmire-et-al-1998}
{Whitmire}, D.~P., {Matese}, J.~J., {Criswell}, L., \& {Mikkola}, S. 1998,
  \icarus, 132, 196

\bibitem[{{Williams} \& {Pollard}(2002)}]{williams-pollard-2002}
{Williams}, D.~M., \& {Pollard}, D. 2002, International Journal of
  Astrobiology, 1, 61

\end{thebibliography}

\onecolumn

\appendix

\section{Insolation Variance}\label{sec:var}
The insolation variance a planet receives on an S-Type orbit in a binary star system was defined in section \ref{subsec:EHZ} as:
\begin{equation}
\sigma^2=\langle S_{tot}^2\rangle_t - \langle S_{tot}\rangle_t^2 \label{eq:app1}
\end{equation}
Considering the linearity of the expectation value operator, the term $\langle S_{tot}\rangle^2$ can be decomposed in\footnote{We drop the subscript $t$ on the averages, 
as there is no danger of misinterpretation.}:
\begin{equation}
\langle S_{tot}\rangle^2= \langle S_1 \rangle^2+2\langle S_1\rangle \langle S_2\rangle + \langle S_2 \rangle^2
\end{equation}
Averages for insolation values from both stars have been derived in section \ref{sec:ana} already, and are therefore not repeated here.
Instead, we will develop expressions for the first term on the right hand side of equation (\ref{eq:app1}).
\begin{equation}
\langle S_{tot}^2\rangle= \langle S_1^2 \rangle+2\langle S_1 S_2\rangle + \langle S_2^2 \rangle
\end{equation}
Using equivalence radii $r$ and $R$ for the planet and the secondary respectively, which were introduced in section \ref{sec:ana}, we get:
\begin{eqnarray}
\langle S_1^2 \rangle &=& \frac{1}{P}\int_0^P \frac{L_1^2}{\delta^4(t)}dt \quad = \quad  \frac{L_1^2}{2\pi r^4}\int_0^{2\pi} dM \quad = \quad \frac{L_1^2}{r^4}  \quad = \quad\frac{L_1^2}{a_p^4(1-\langle e_p^2 \rangle)^4} \label{eq:s1sq} \\
\langle S_1S_2 \rangle &=& \frac{1}{2\pi}\int_0^{2\pi} \frac{L_1}{r^2}\frac{L_2}{R^2+r^2- R r \cos(\phi)} d\phi \quad = \quad \frac{L_1L_2}{R^2r^2-r^4}  \label{eq:s1s2} \\
\langle S_2^2\rangle &=& \frac{1}{2\pi}\int_0^{2\pi} \left(\frac{L_2}{R^2+r^2- R r \cos(\phi)}\right)^2 d\phi \quad = \quad -\frac{L_2^2(R^2+r^2)}{(r^2-R^2)^3}  \label{eq:s2sq} 
\end{eqnarray}
where $\delta(t)$ is again the time dependent distance of the planet to its host star, and $a_p$ and $e_p$ the planetary semi-major axis and eccentricity respectively.
Given the circular nature of the orbital equivalence approximations we have applied, the results are bound to underestimate the true variances. 
Since the radiative contribution of the primary dominates the planetary insolation for S-Type I systems and is at least as improtant as the secondary's insolation for the S-Type II systems investigated, 
we will use stronger estimates for $\langle S_1^2\rangle_t$ than given in relation (\ref{eq:s1sq}):
\begin{eqnarray}
\langle S_1^2\rangle&=&\frac{1}{P}\int_0^P \frac{L_1^2}{\delta^4(t)}dt \quad = \quad \frac{L_1^2 n}{2\pi h}\int_0^{2\pi} \frac{1}{\delta^2}df \nonumber \\
                    &=& \frac{L_1^2 n}{2\pi h}\int_0^{2\pi} \left(\frac{1+\langle e_p\rangle \cos(f)}{a_p(1-\langle e_p^2 \rangle)}\right)^2 df \nonumber \\
                    &=& \frac{L_1^2 (1+\langle e_p^2 \rangle/2)}{a_p^4 (1-\langle e_p^2 \rangle)^{5/2}}  \quad = \quad  
                    \frac{L_1^2}{r^4}\left(1-\langle e_p^2 \rangle \right)^{3/2} \left(1+\frac{\langle e_p^2 \rangle}{2}\right) \label{eq:s1sq2}
\end{eqnarray}
As one can see, the difference between relations (\ref{eq:s1sq}) and (\ref{eq:s1sq2}) is negligible for small injected planetary eccentricities, 
but its contribution becomes important if the injected eccentricities grow.
 Combining expressions (\ref{eq:s1sq2}), (\ref{eq:s1s2}) and (\ref{eq:s2sq}) with the respective terms of $\langle S_{tot} \rangle_t^2$ produces the desired variance:
\begin{eqnarray}
%\sigma^2 &=&\frac{L_1^2\left(1-\frac{3}{2} e^2+\frac{1}{2}e^6- (1-e^2)^{1/2}\right) }{a^4 \left(1-e^2\right)^{9/2}}-\frac{2 L_2^2 a^2 \left(-1+e^2\right)^2 }{\left(a^2 \left(-1+e^2\right)^2-R^2\right)^3}
%\sigma^2 &=&\frac{L_1^2}{r^4}\left(\sqrt{1-\langle e_p^2\rangle}\left(1-\frac{\langle e_p^2\rangle}{2}-\frac{\langle e_p^2\rangle^2}{2}\right)-1\right)-\frac{2 L_2^2 r^2}{\left(r^2-R^2\right)^3}
\sigma^2 &=&\frac{L_1^2}{r^4}\left(-1+3\langle e_p^2\rangle-3\langle e_p^2\rangle^2+\langle e_p^2\rangle^3+\sqrt{1-\langle e_p^2\rangle}\left(1-\frac{\langle e_p^2\rangle}{2} - \frac{\langle e_p^2\rangle^2}{2} \right)\right) \nonumber \\ 
            &&-\frac{2L_1L_2}{r^4-r^2R^2}\left(1-\sqrt{1-\langle e_p^2\rangle}\left(1+\langle e_p^2\rangle\right)\right)-
            \frac{2L_2^2r^2}{(r^2-R^2)^3}
\end{eqnarray}

\section{Maximum Planetary Eccentricity}\label{sec:emax}
The maximum possible eccentricity a terrestrial planet's orbit can acquire in an S-Type setup is composed of 
\begin{equation}
e^{max}_p=e^{sp}_p+e^{sec}_p
 \end{equation}
where $e^{sp}_p$ denotes amplitude of short-period terms and $e^{sec}_p$ the planetary eccentricity's secular amplitude. 
Using expressions derived in \citet{georgakarakos-2003} via the Laplace-Runge-Lenz vector, we can estimate the maximum short-period contributions for the planet's eccentricity by
taking only terms including the dominant frequencies into account. The amplitude of the secular part of the planet's eccentricity is given by $2C/(B-A)$ \citep{georgakarakos-2003} resulting in the following expressions:  
\begin{eqnarray}
e^{sp}_p&=&\alpha \left(\frac{15}{64}\frac{\beta}{X^{5/3}}\frac{(4+11 e_b^2)}{(1-e_b^2)^{5/2}}+\frac{11}{4}\frac{1}{X^{2}}\frac{(1+e_b)^3}{(1-e_b^2)^3}+\frac{3}{4}\frac{1}{X^{3}}\frac{(1+e_b)^4(6+11 e_b)}{(1-e_b^2)^{9/2}}\right) \\
e^{sec}_p&=& e_b\beta \left(\frac{5}{4}\frac{\alpha}{X^{1/3}}\frac{3+2e_b^2}{(1-e_b^2)^{1/2}} -\frac{2}{5}\gamma X^{1/3}(1-e_b^2)^{1/2} +\frac{2}{5} X^{2/3}(1-e_b^2)\right)^{-1} 
\end{eqnarray}
The mass parameters $\alpha,\beta$ and $\gamma$ are defined as:
\begin{displaymath}
 \alpha=\frac{m_2}{M}\qquad \qquad \beta=\frac{m_1-m_p}{(m_1+m_p)^{2/3} M^{1/3}} \qquad \qquad \gamma=\frac{m_1 m_p M^{1/3}}{m_2(m_1+m_p)^{4/3}}
\end{displaymath}
 $m_p$ being the planetary mass and $m_1, m_2$ the stellar masses of primary and secondary respectively.\\ \mbox{$M=m_1+m_2+m_p$} is the total mass of the system.  
 Finally, $X$ denotes the secondary to planet period ratio:
\begin{displaymath}
 X=\frac{P_b}{P_p}=\left(\frac{m_1+m_p}{M}\right)^{1/2}\left(\frac{a_b}{a_p}\right)^{3/2}
\end{displaymath}

\begin{figure}
%\epsscale{.80}
\begin{tabular}{ll}
 \includegraphics[angle=-90, scale=0.34]{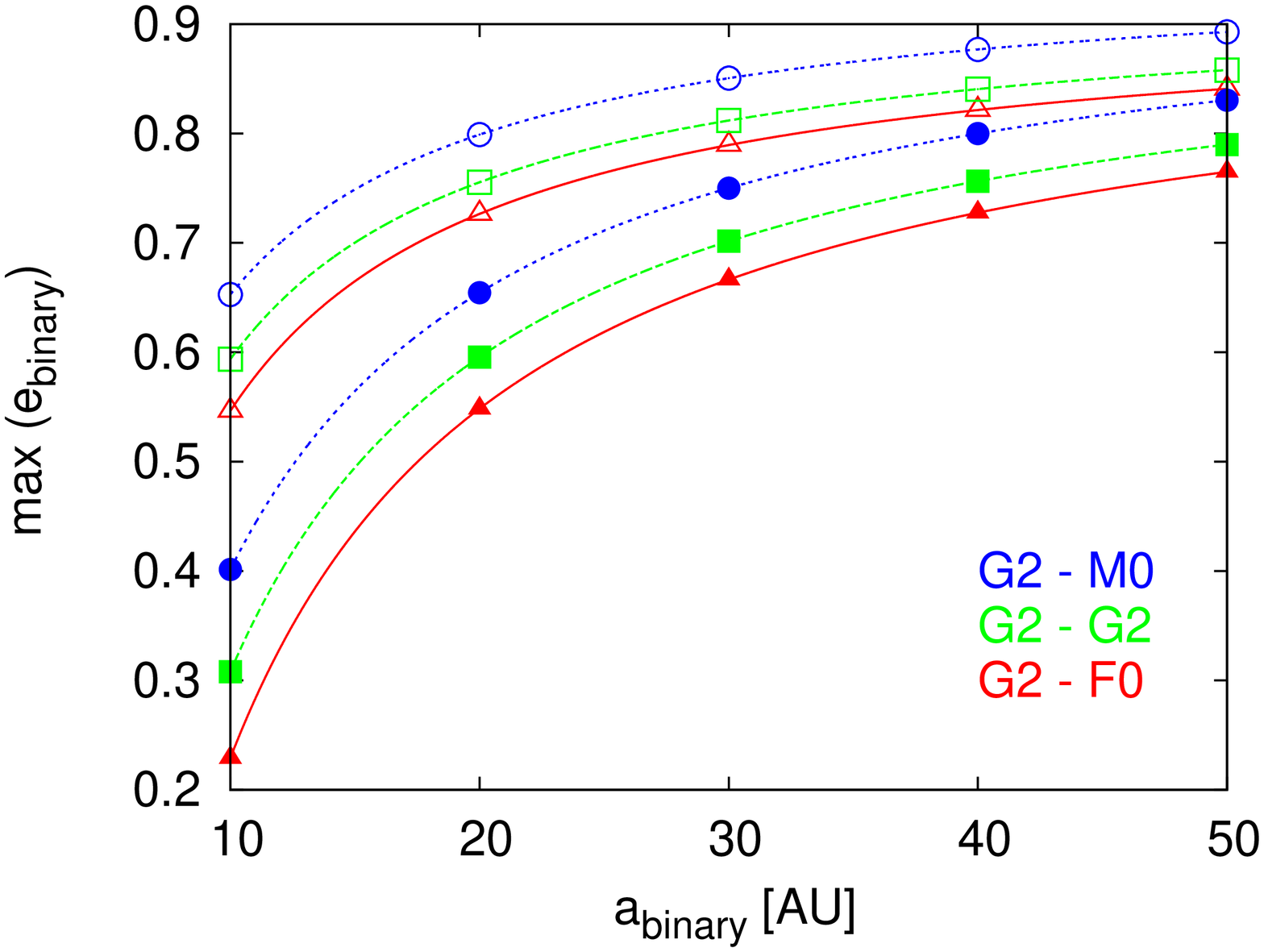}  
 \includegraphics[angle=-90, scale=0.34]{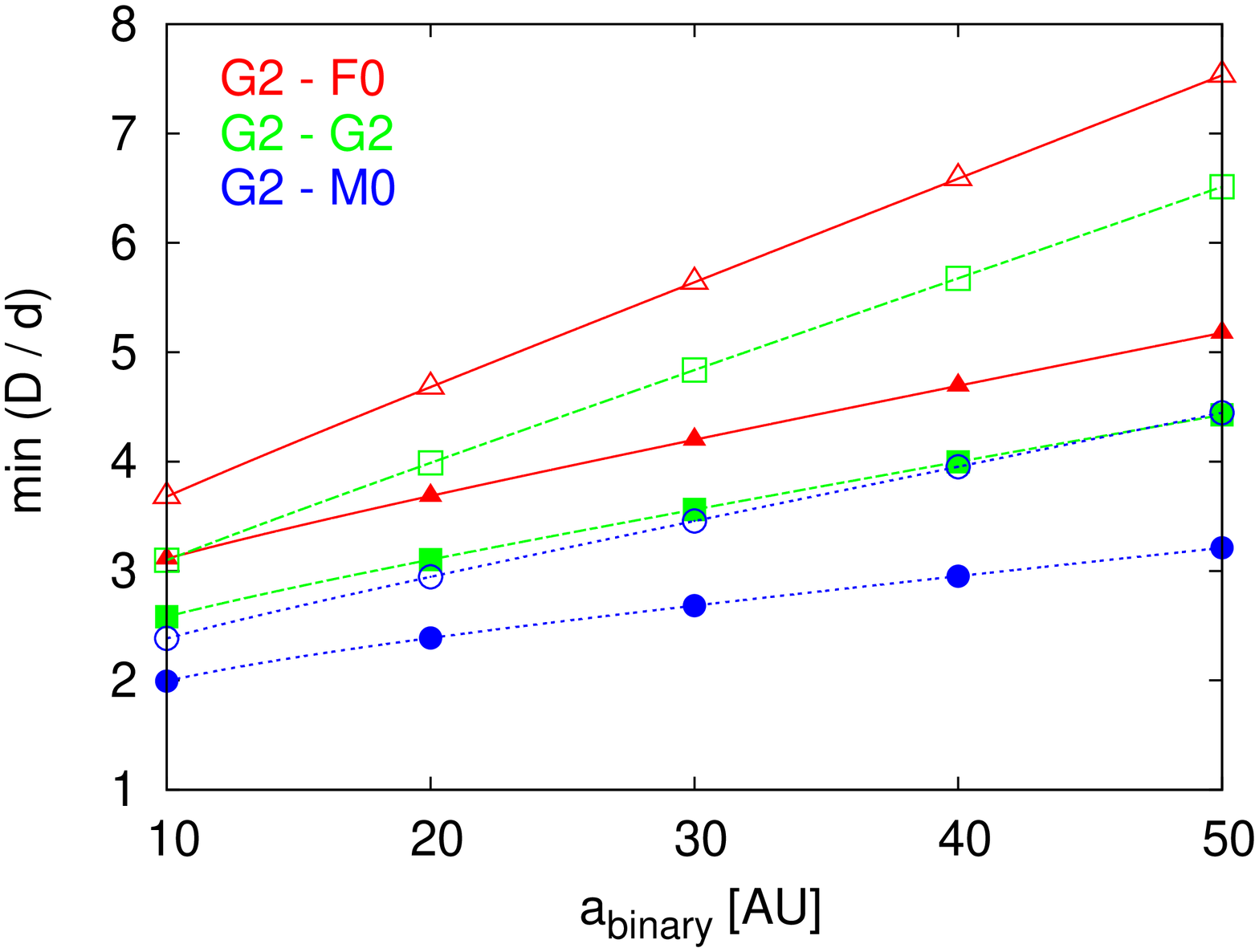}\\ 
\includegraphics[angle=-90, scale=0.34]{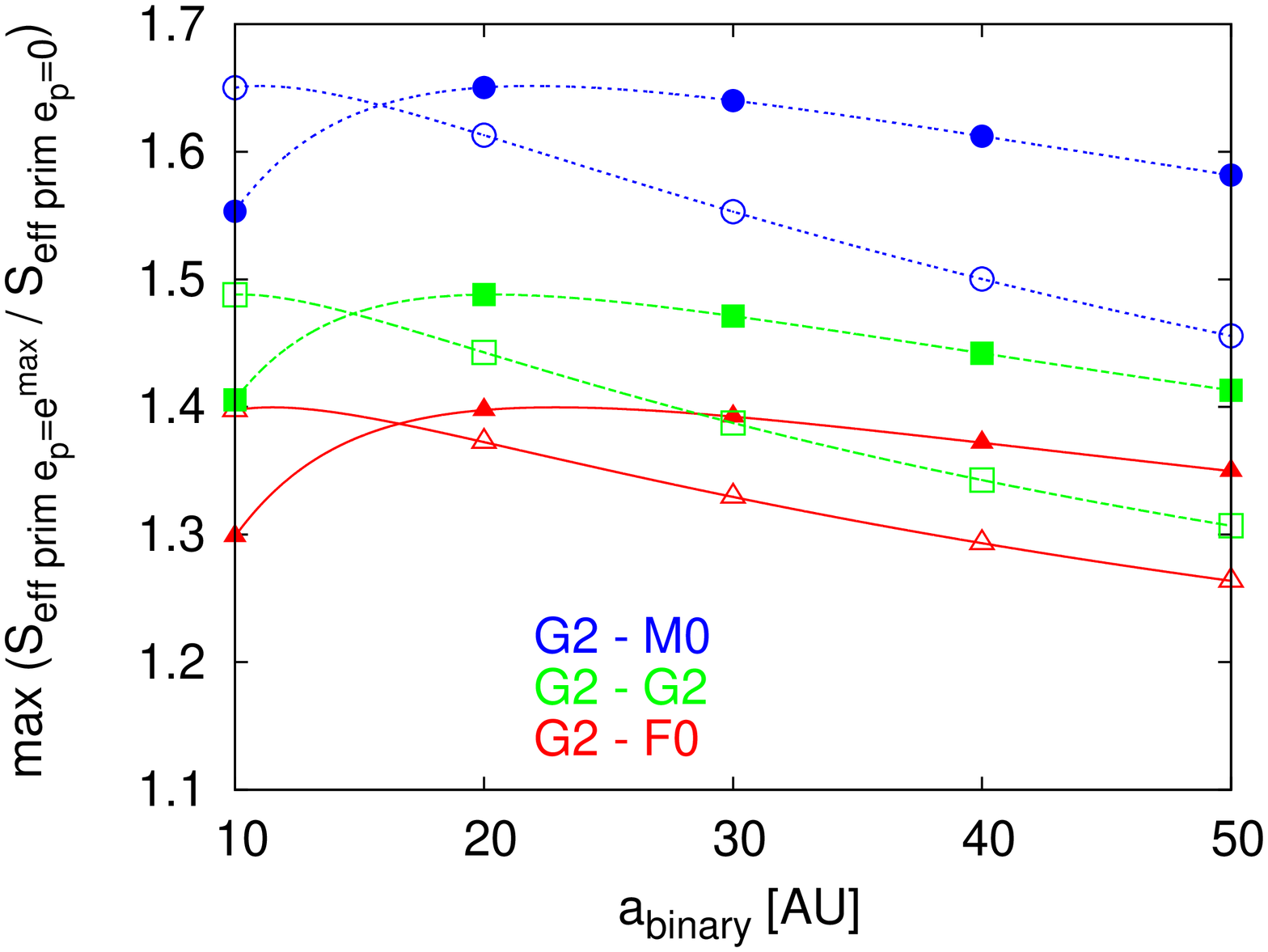} 
\includegraphics[angle=-90, scale=0.34]{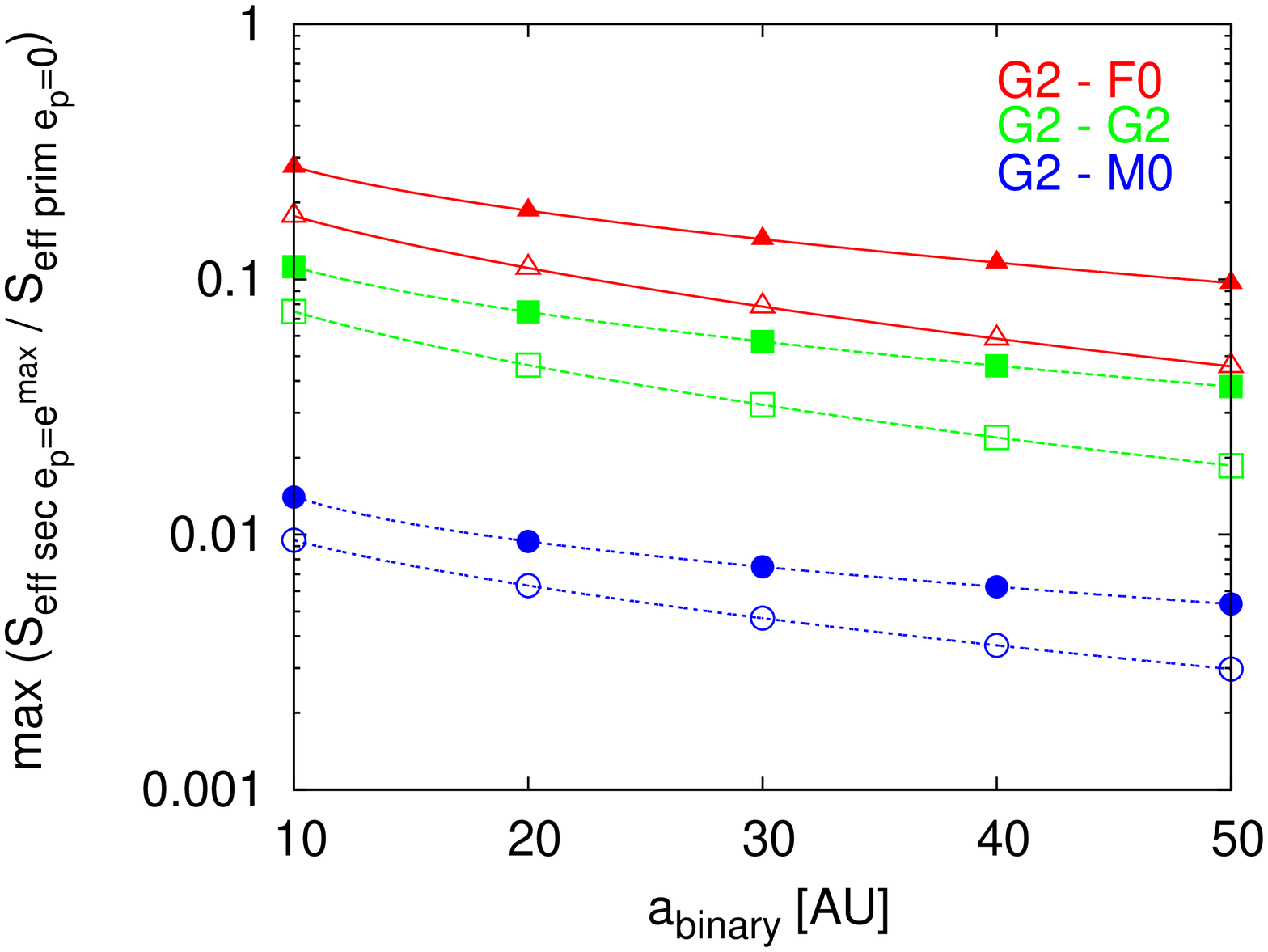} 
\end{tabular}
\caption{
\textit{top left:} Highest possible binary eccentricity as a function of the secondary's semi-major axis, if a terrestrial planet was to remain dynamically stable on orbits with semi-major axes corresponding to the inner (runaway greenhouse, open symbols) or outer (maximum greenhouse, full symbols) boundaries of the HZ as 
defined by \citet{kasting-et-al-1993}. The curves represent quadratic fits of the FLI stability data presented in Tab. \ref{tab1}. 
Three different S-Type stellar configurations are shown: G2-F0 ($\triangle, \blacktriangle$), G2-G2 ($\square, \blacksquare$) and  G2-M0 ($\circ, \bullet$). 
\mbox{\textit{top right:}} Minimum distance between planet and secondary ($D$) permitting planetary orbital stability in units of planetary aphelion distances from the primary ($d=a_p (1+e^{max}_p)$). 
The planet's eccentricity was estimated following \citet{georgakarakos-2005}.  
\mbox{\textit{bottom left:}} Here, the increase of the primary's effective insolation onto a terrestrial planet with injected eccentricity is compared to a single-star setup where the planet remains on a circular orbit. 
The planet is considered to be in periastron position ($a_p (1-e_p^{max})$).    
\mbox{\textit{bottom right:}} The secondaries' maximum radiative contributions to planetary insolation are presented - the planet is in apocenter position with regard to the primary ($d$), and the secondary is at pericenter. 
Once again, the results are normalized with regard to values which a single-star configuration with a terrestrial planet on a constant circular orbit would exhibit.
One can see that the primary's radiative influence dominates in S-Type I systems, whereas for close S-Type II configurations the secondary's contribution is almost equally important.  \label{fig2}}
\end{figure}

\begin{figure}
\begin{tabular}{ll}
\includegraphics[angle=-90, scale=0.34]{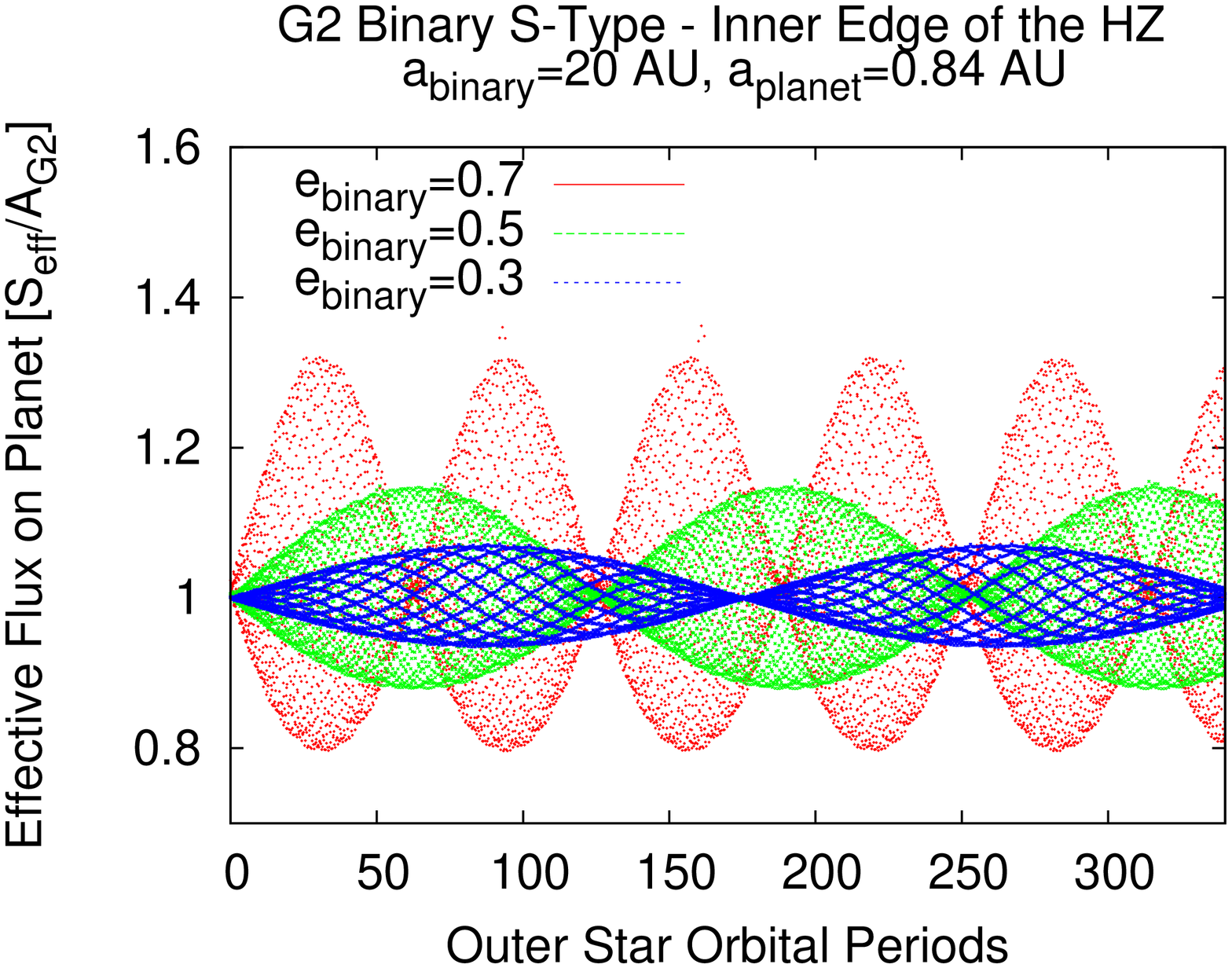}
\includegraphics[angle=-90, scale=0.34]{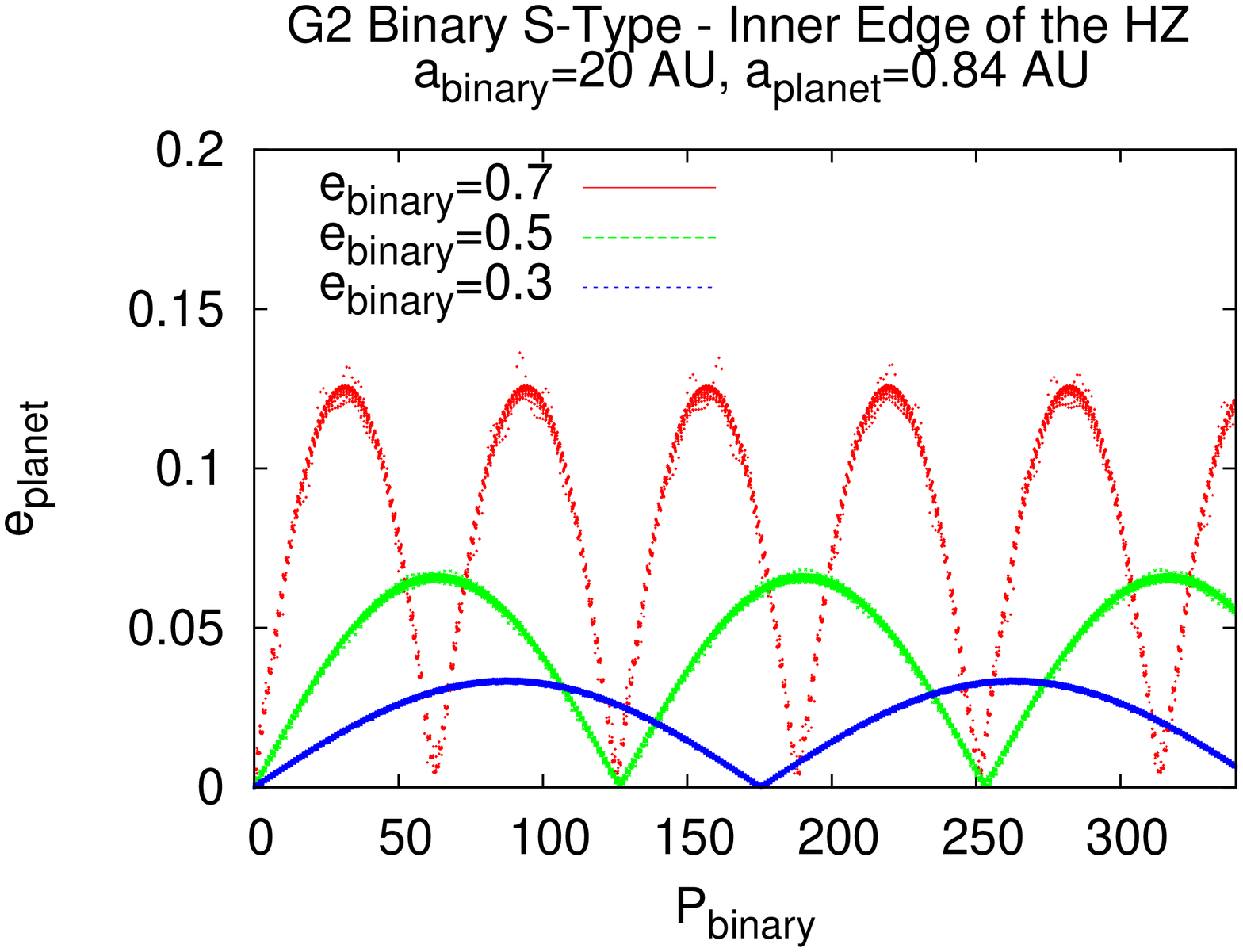}
\end{tabular}
\caption{Evolution of insolation onto an Earth-like planet in a G2 - G2 binary star system. 
The oscillations (\textit{left}) are due to the injected changes in the planet's eccentricity (\textit{right}) caused by the gravitational perturbations of the secondary.   
See the electronic edition of the Journal for a color version 
of this figure.\label{fig3}}
\end{figure}

\begin{figure}
\begin{tabular}{cc}
Analytics & Simulation\\
\includegraphics[angle=-90, scale=0.32]{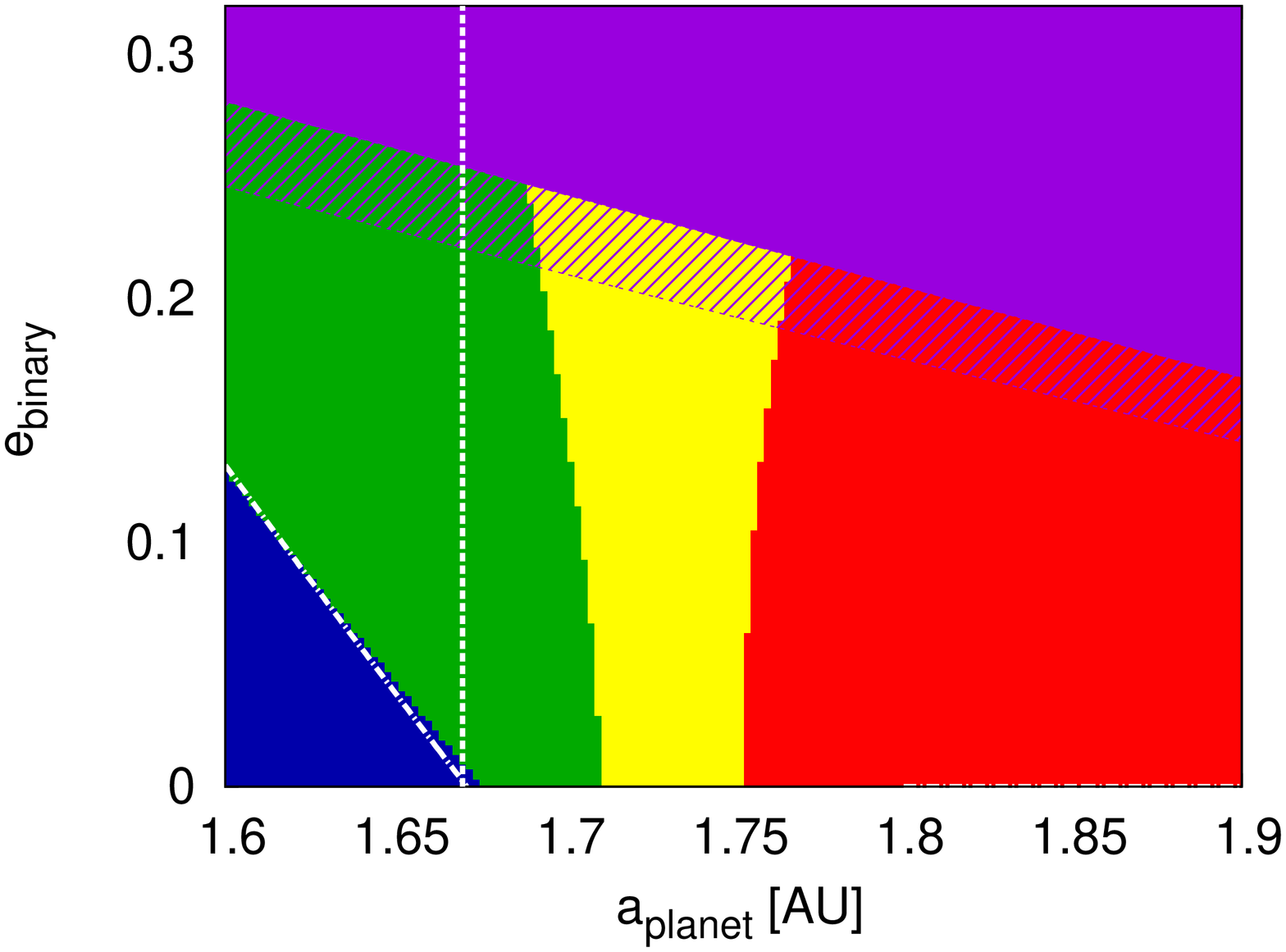} &
 \includegraphics[angle=-90, scale=0.32]{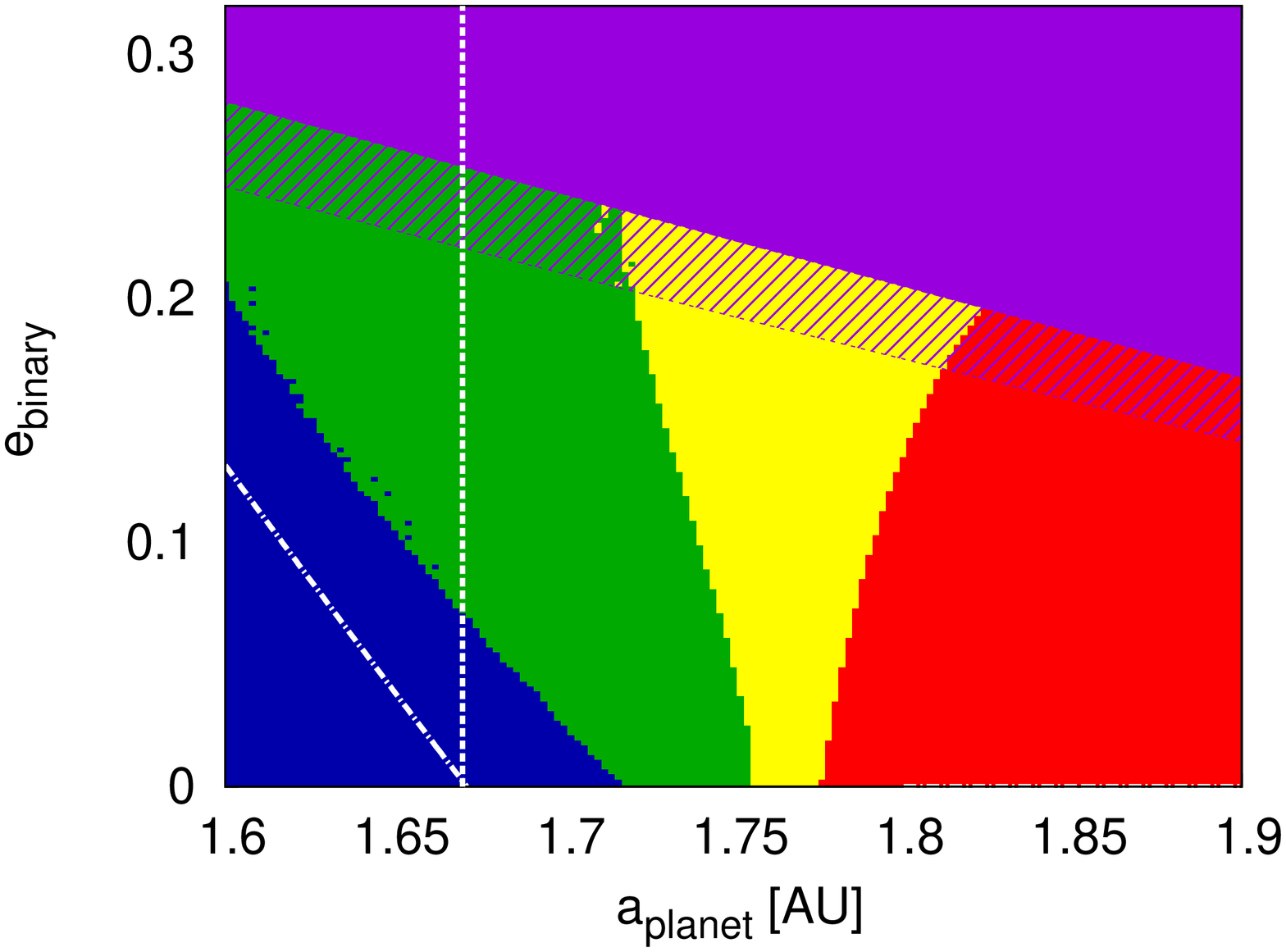}\\
\end{tabular}
\caption{Classification of HZs in a G2-F0 binary star system of S-Type II with a semi-major axis $a_{b} = 10\;AU$.{ \it left:} 
Zoom on the outer limit of the classical HZ (dashed line) close to the instability region (purple). The results were gained using 
analytic estimates presented in section~\ref{sec:ana}. 
Black (blue online) denotes the PHZ, dark gray (green online) the EHZ, light gray (yellow online) 
the AHZ and white (red online) indicates that the planet is not habitable. 
The gray striped area (purple online) corresponds to the dynamically unstable region (HW99), the striped extension shows the onset of dynamical chaos (PLD02).
{\it right:} The numerical simulation results for the same configuration. 
The HZ limits extend beyond the values defined in \citet{kasting-et-al-1993}. However, the
white dashed-dotted line corresponds to the semi-analytic estimates of the PHZ using numerically derived values for $e_p^{max}$. 
The semi-analytic results agree with the fully analytic
estimates. This may indicate shortcomings of the entirely numerical approach to identify PHZ boundaries.
 The resolution of these calculations is $\Delta a_p=0.002\,AU$,
and $\Delta e_b=0.002$.  
 \label{fig7}}
\end{figure}

\begin{figure}
\begin{tabular}{cc}
Analytics & Simulation \\
\includegraphics[angle=-90, scale=0.32]{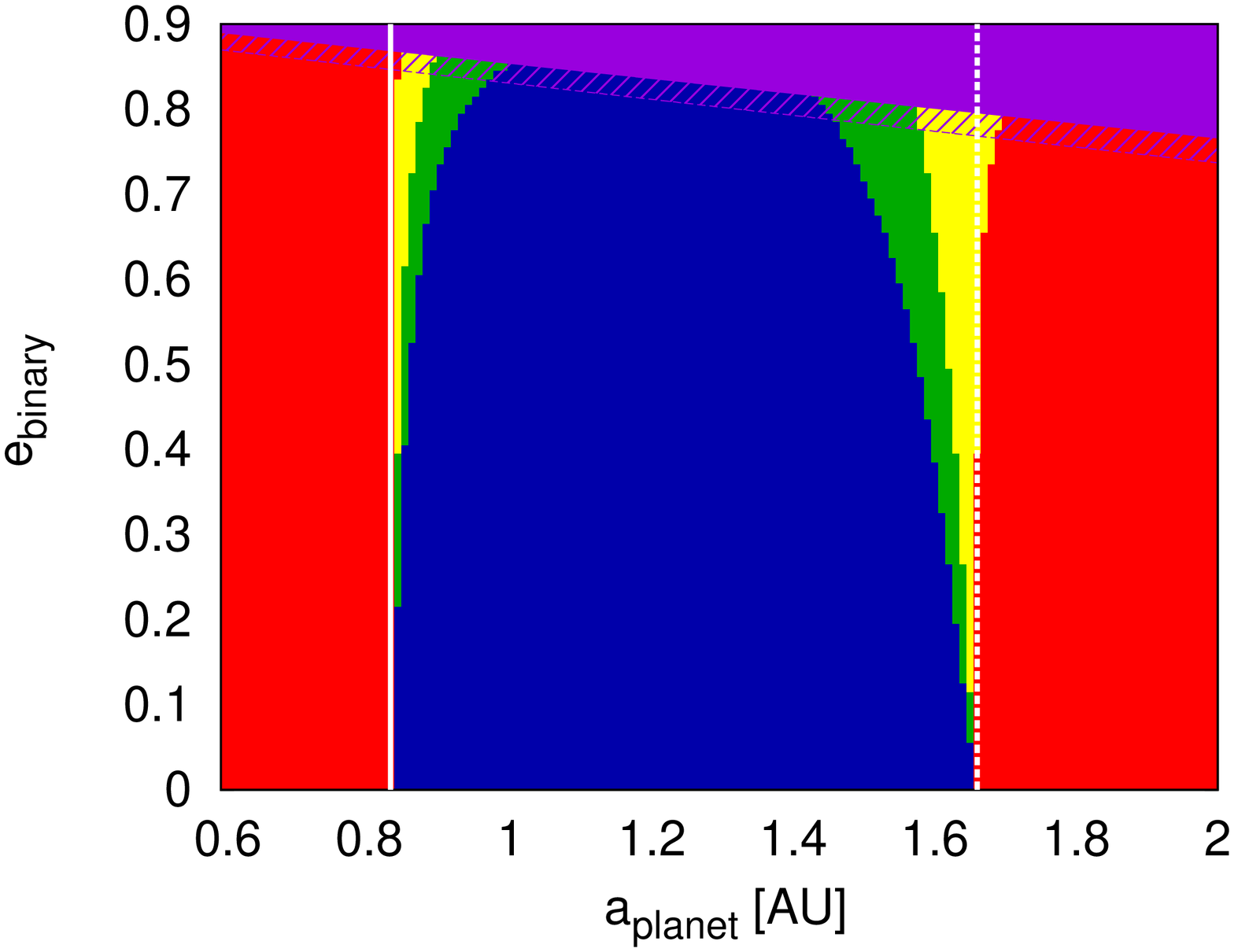} &
 \includegraphics[angle=-90, scale=0.32]{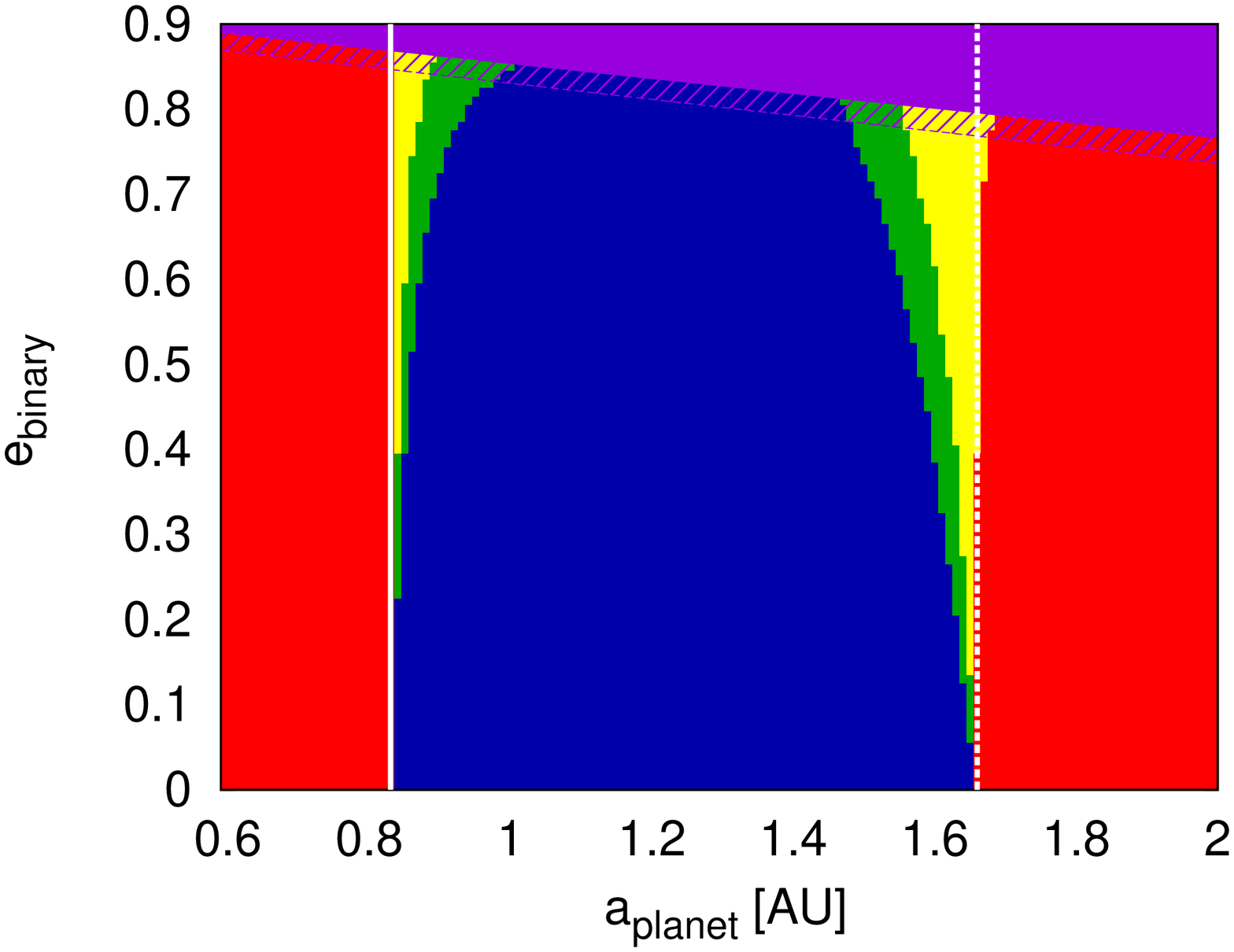}\\
 \includegraphics[angle=-90, scale=0.32]{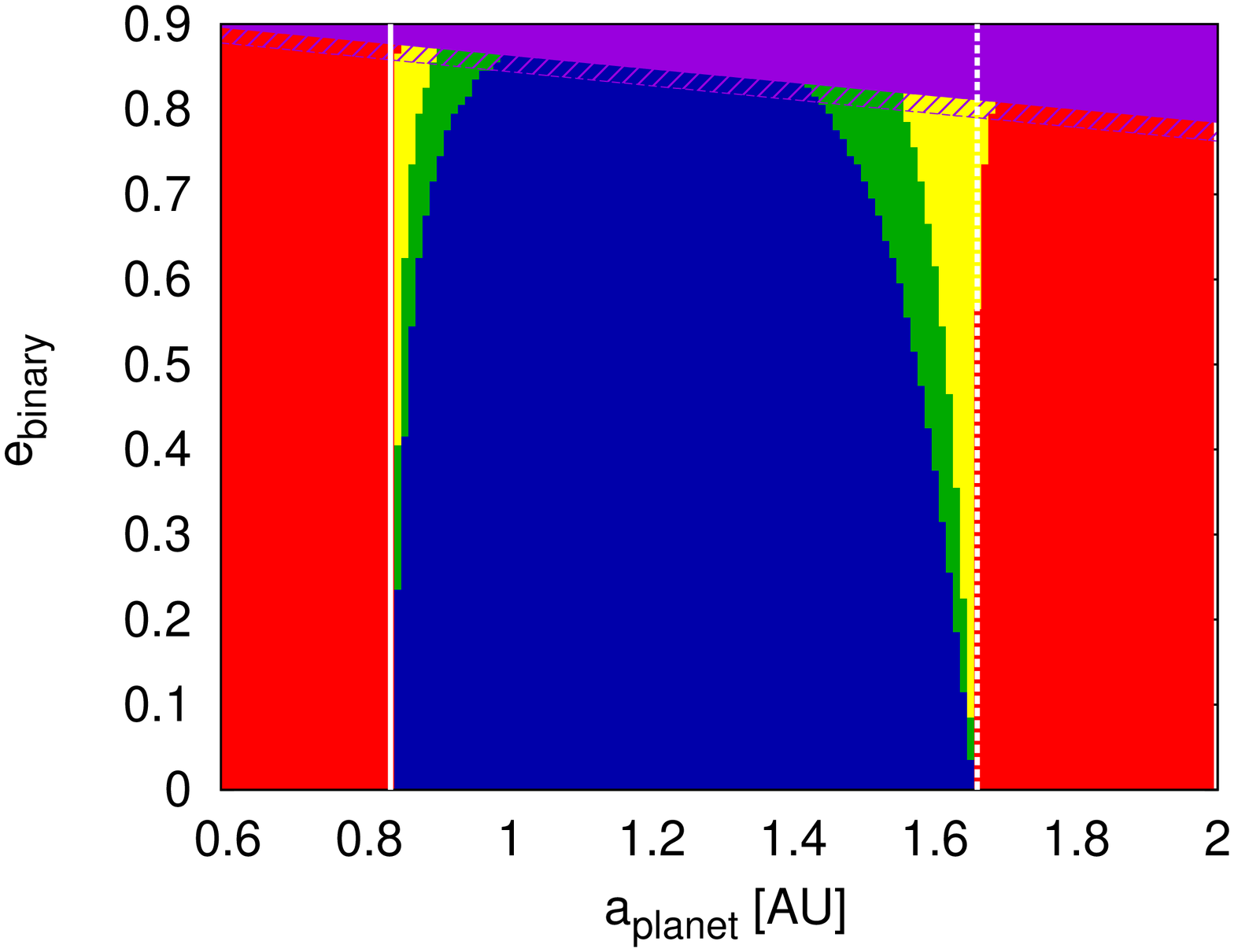}&
\includegraphics[angle=-90, scale=0.32]{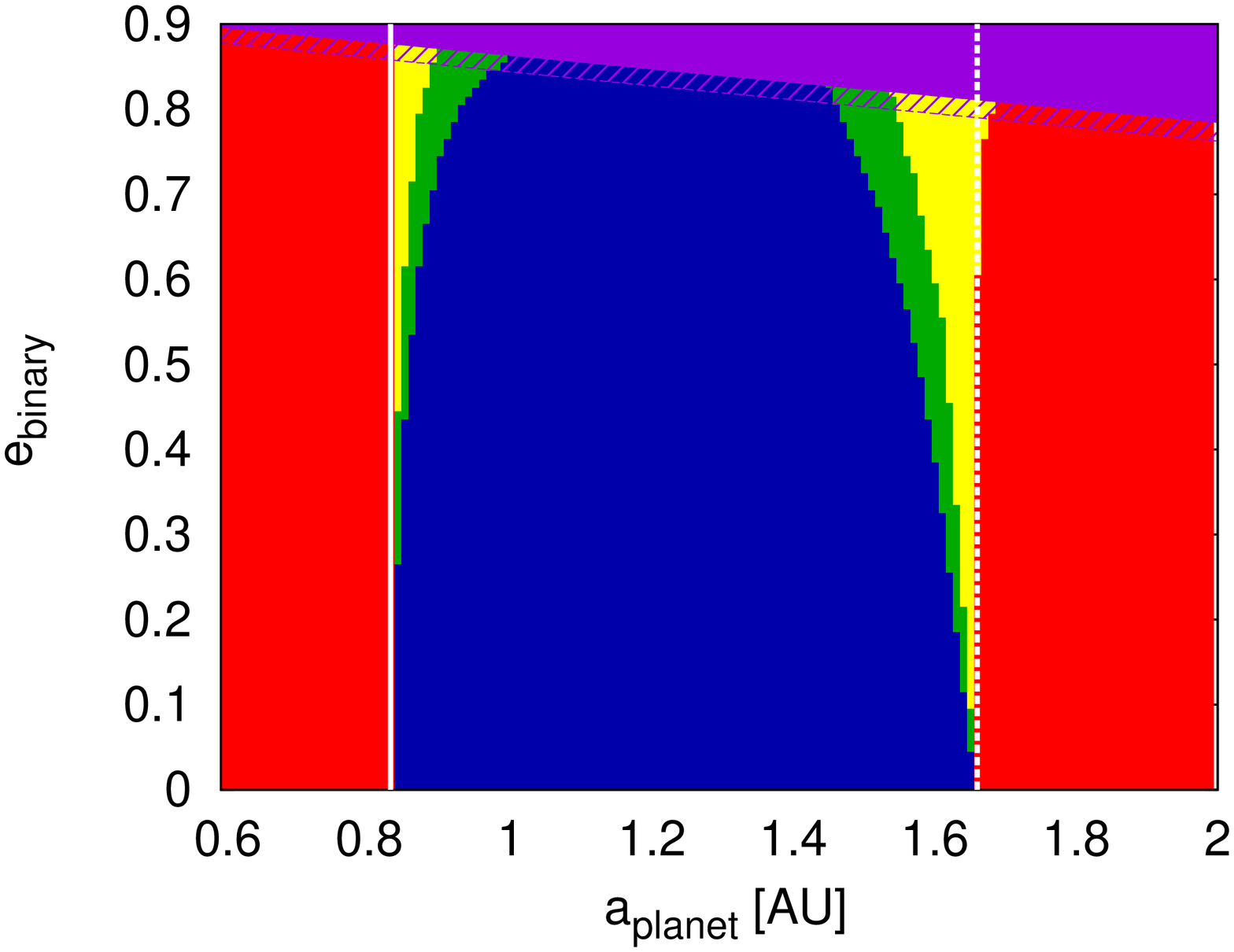}\\
 \includegraphics[angle=-90, scale=0.32]{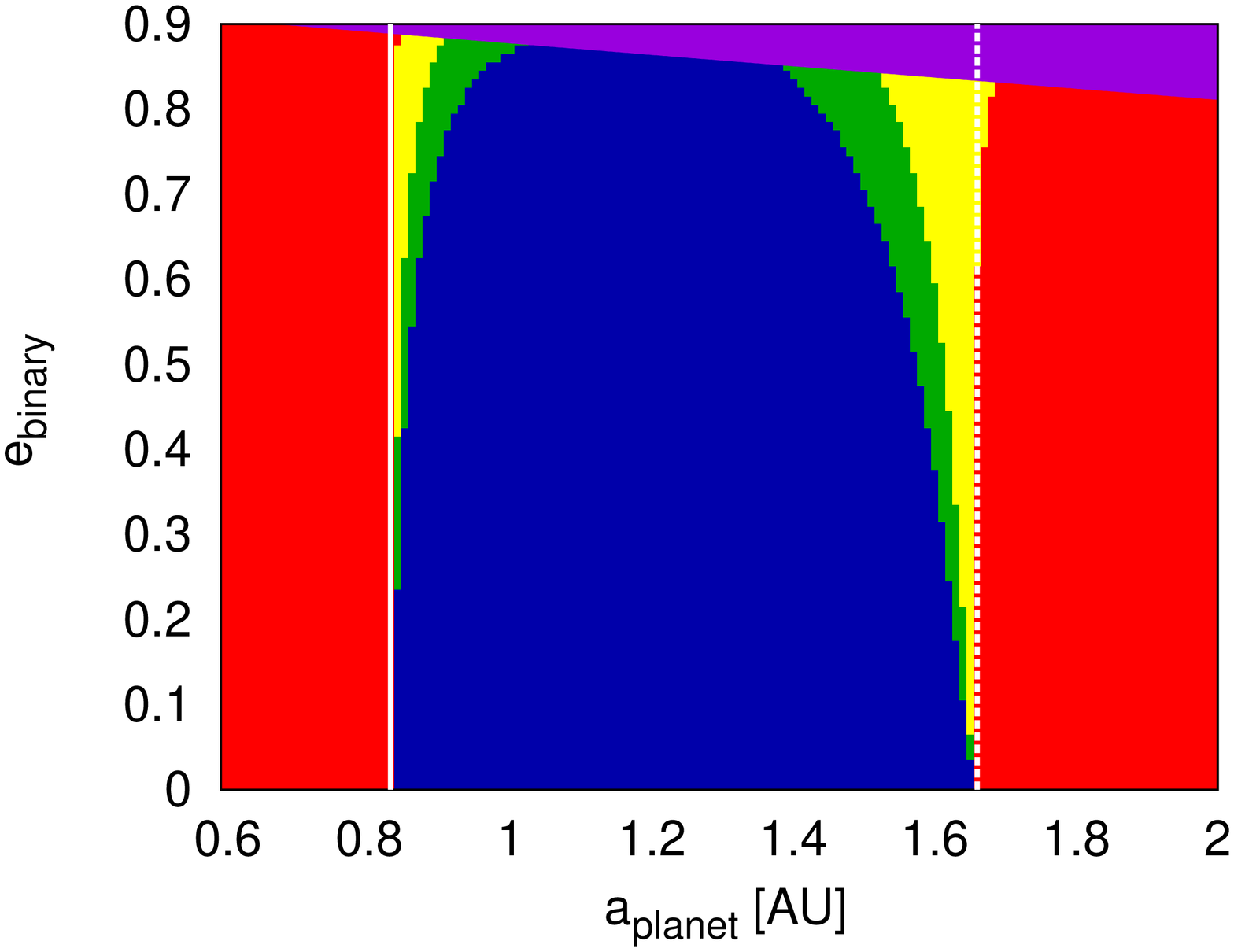} &
\includegraphics[angle=-90, scale=0.32]{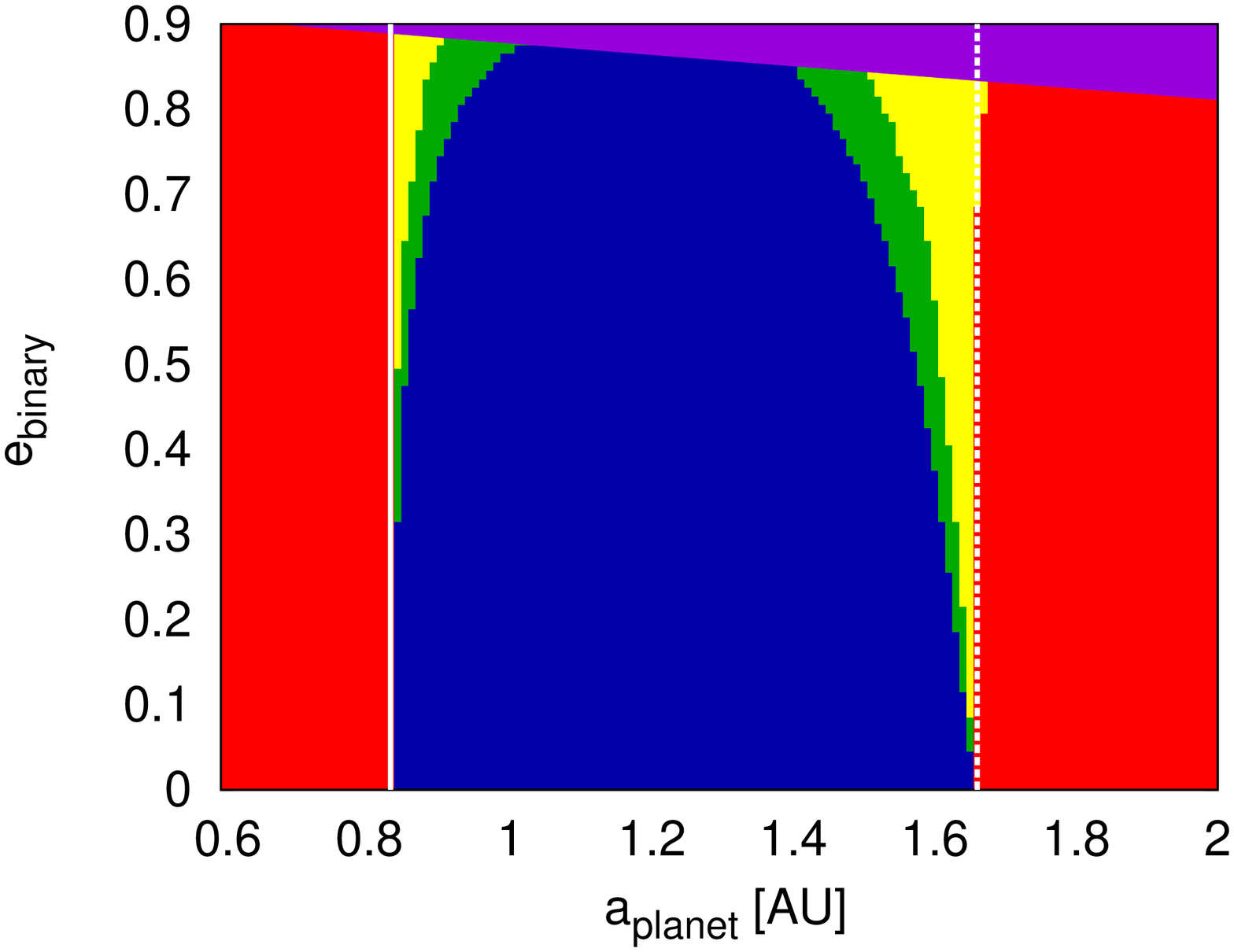}\\
\end{tabular}
\caption{Comparison of HZ classifications in three binary star systems with a secondary's semi-major axis of $a_{b} = 50\;AU$. 
{\it left:} analytical estimates as discussed in section\,\ref{sec:ana},    
{\it right:} reference results gained via numerical integration, with a resolution of $\Delta a_p=0.01\,AU$,
and $\Delta e_b=0.01$.  
The investigated stellar spectral configurations are: {\it top:} G2-F0, {\it mid:} G2-G2, {\it bottom:} G2-M0;
The PHZ is represented in black (blue online), the EHZ is dark gray (green online) , light gray (yellow online) 
indicates the AHZ and white regions (red online) mean that the planet is outside of any defined HZ. 
The gray striped area (purple online) denotes dynamically unstable parameter regions (HW99), whereas the striped extension highlights the onset of dynamical chaos (PLD02), 
see section \ref{sec:stab}. 
The borders of the classical HZ as defined in KWR93 are represented by the vertical solid and dashed lines. \label{fig5}} \end{figure}

\begin{figure}
\begin{tabular}{cc}
Analytics & Simulation \\
\includegraphics[angle=-90, scale=0.32]{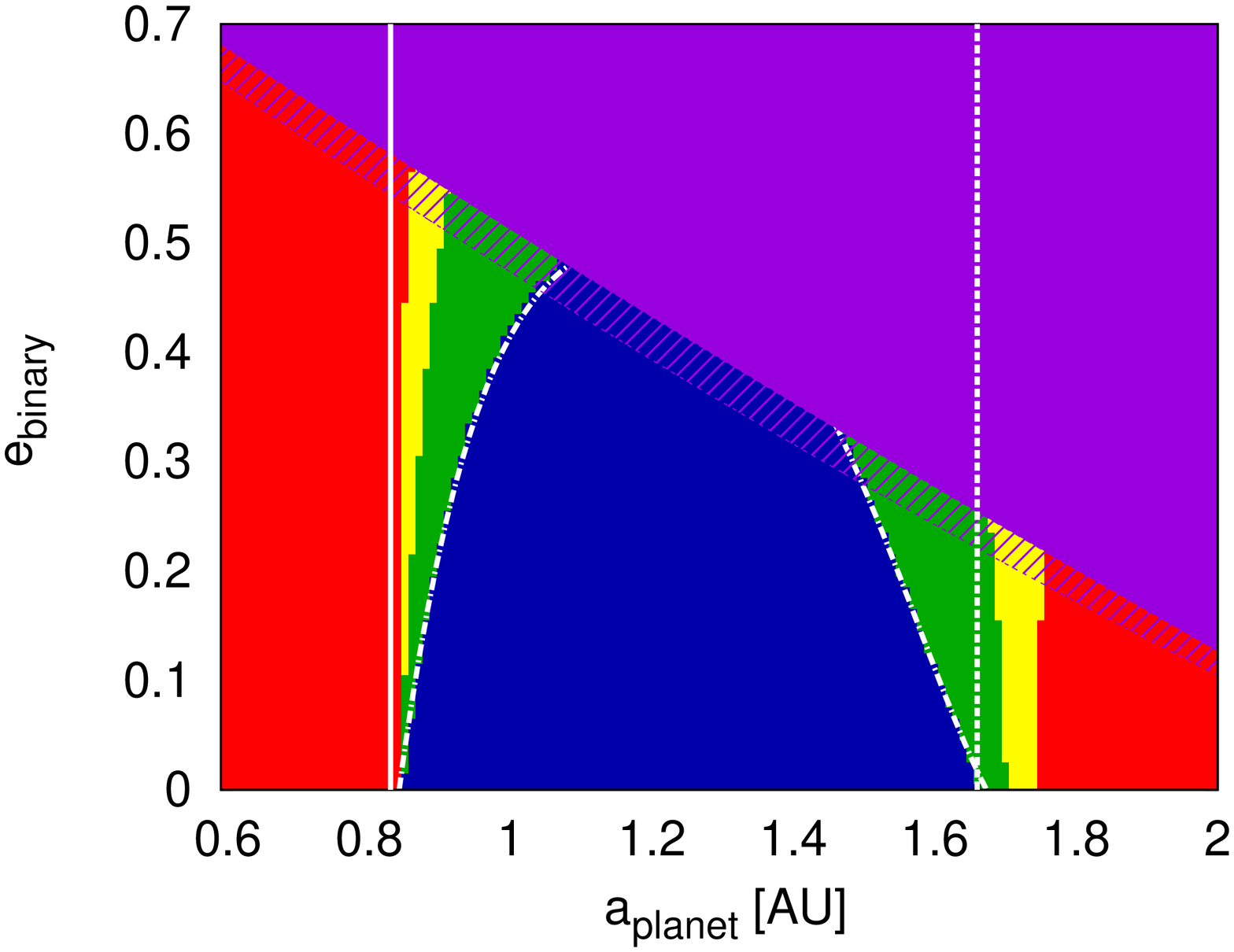} &
 \includegraphics[angle=-90, scale=0.32]{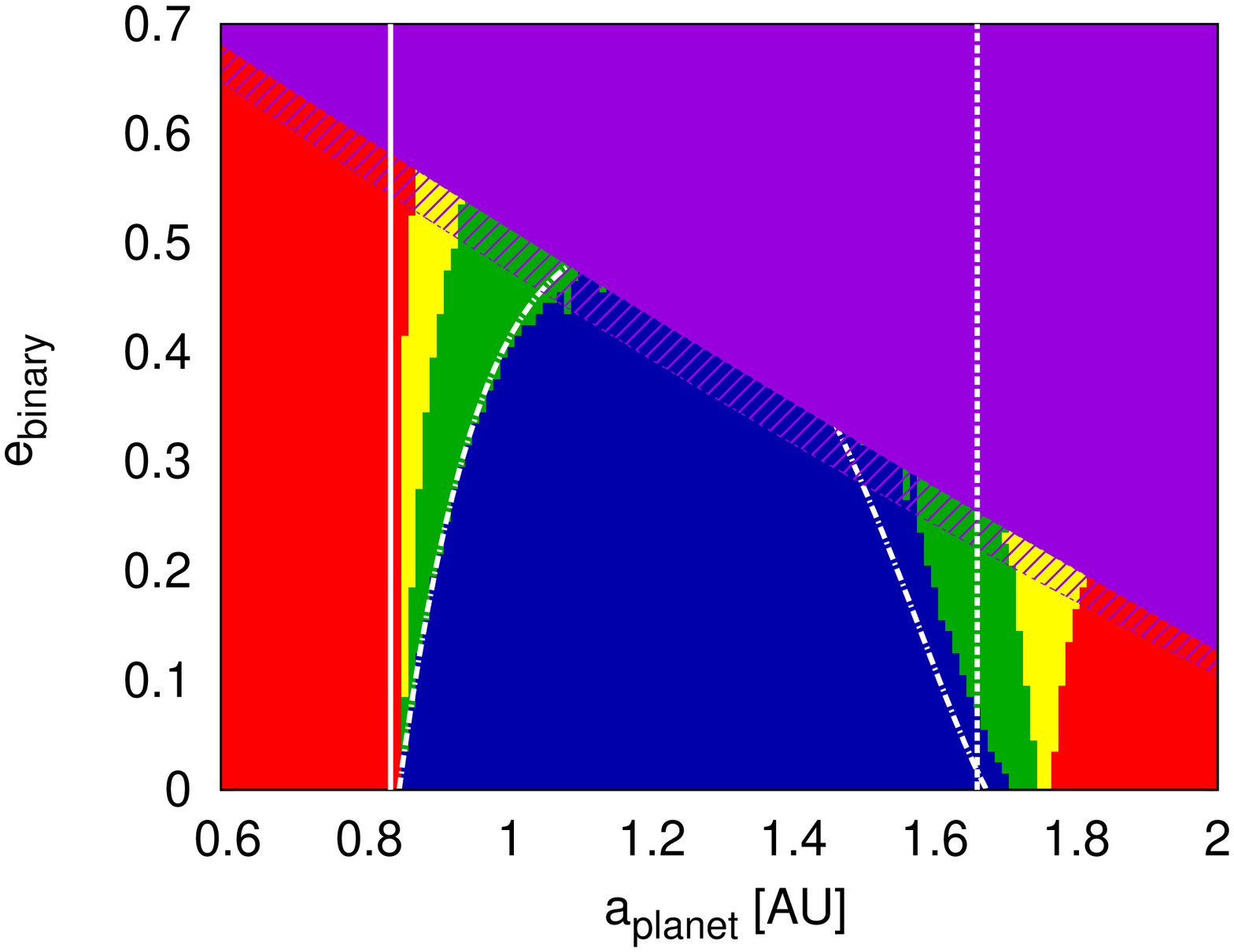}\\
\includegraphics[angle=-90, scale=0.32]{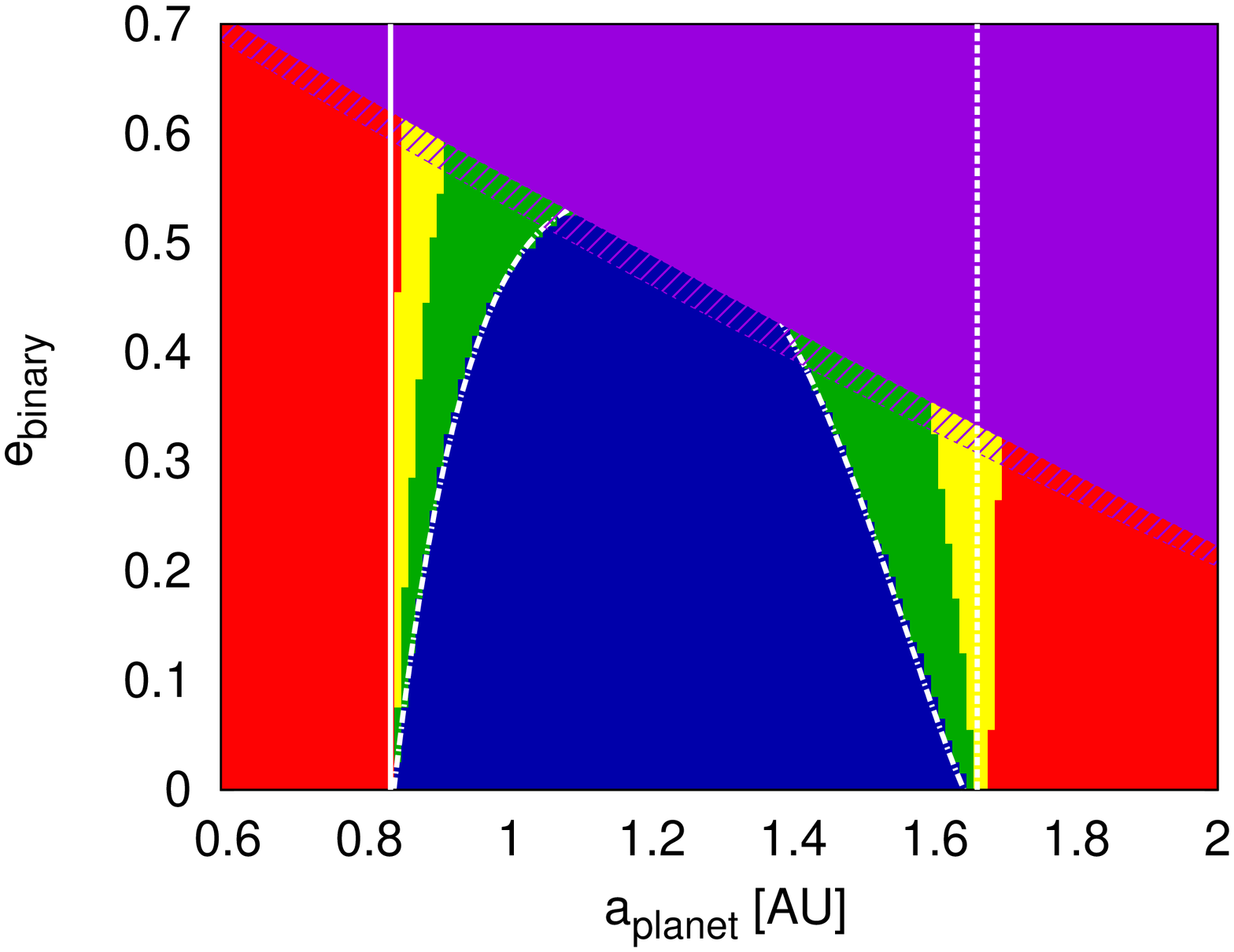} &
 \includegraphics[angle=-90, scale=0.32]{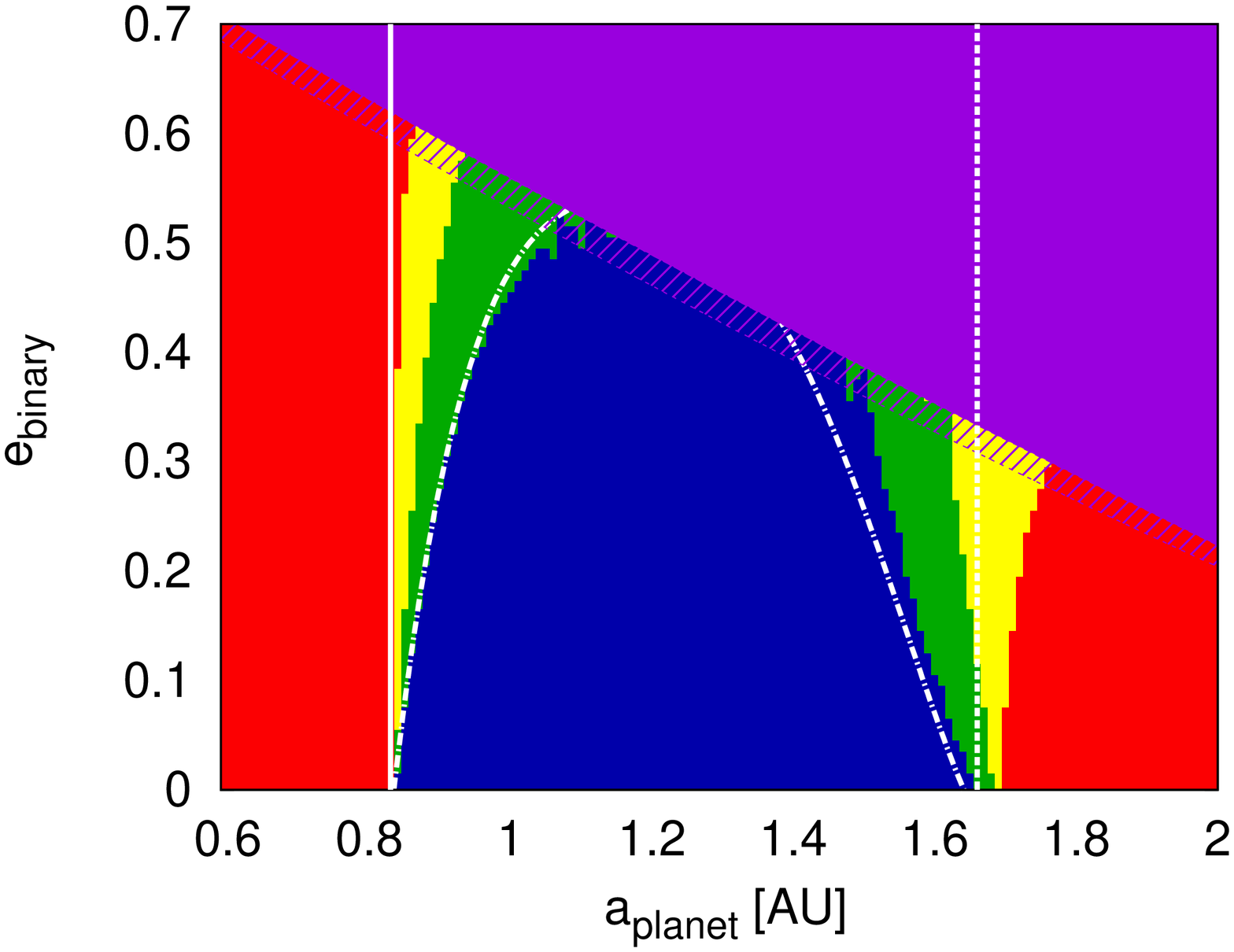}\\
\includegraphics[angle=-90, scale=0.32]{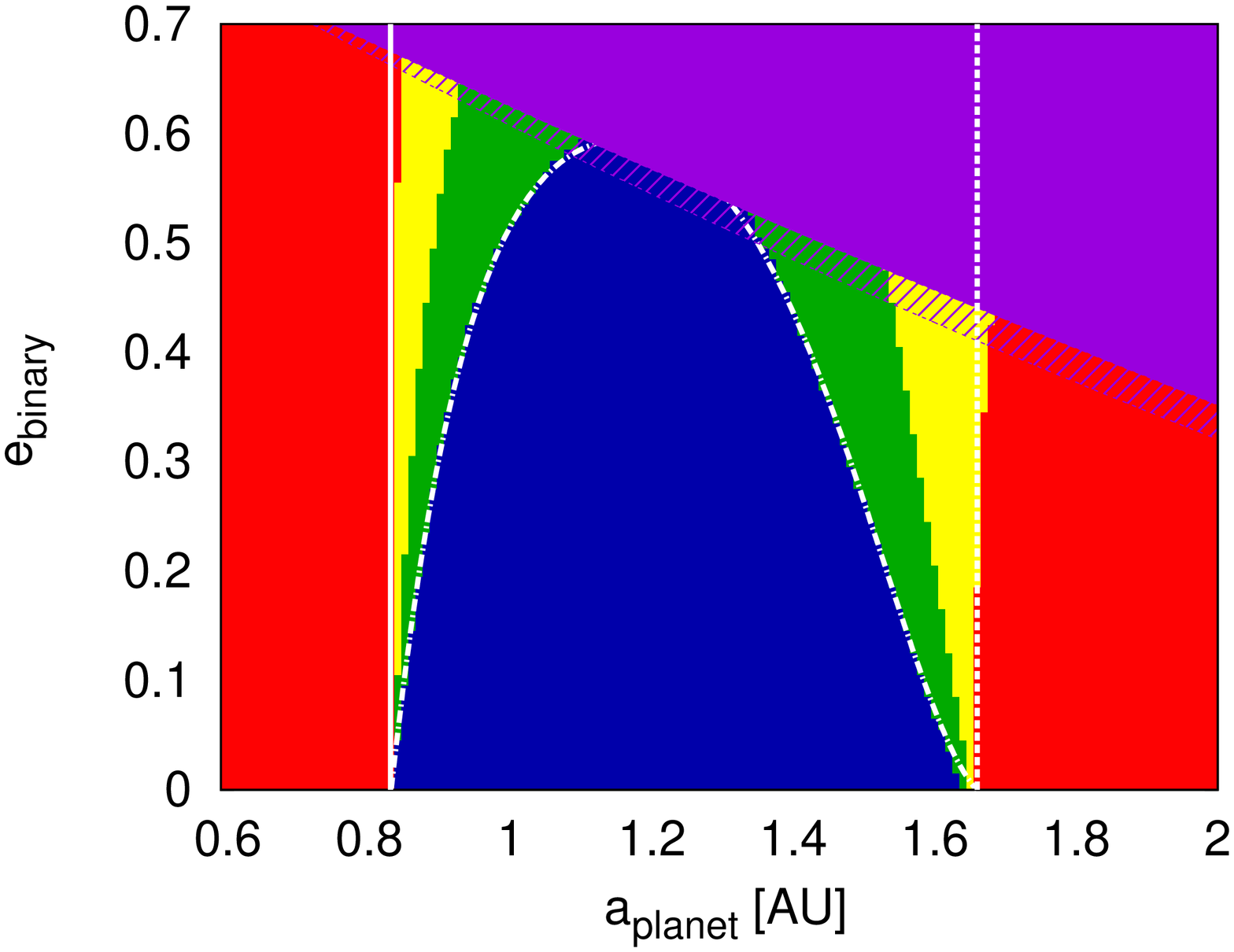} &
 \includegraphics[angle=-90, scale=0.32]{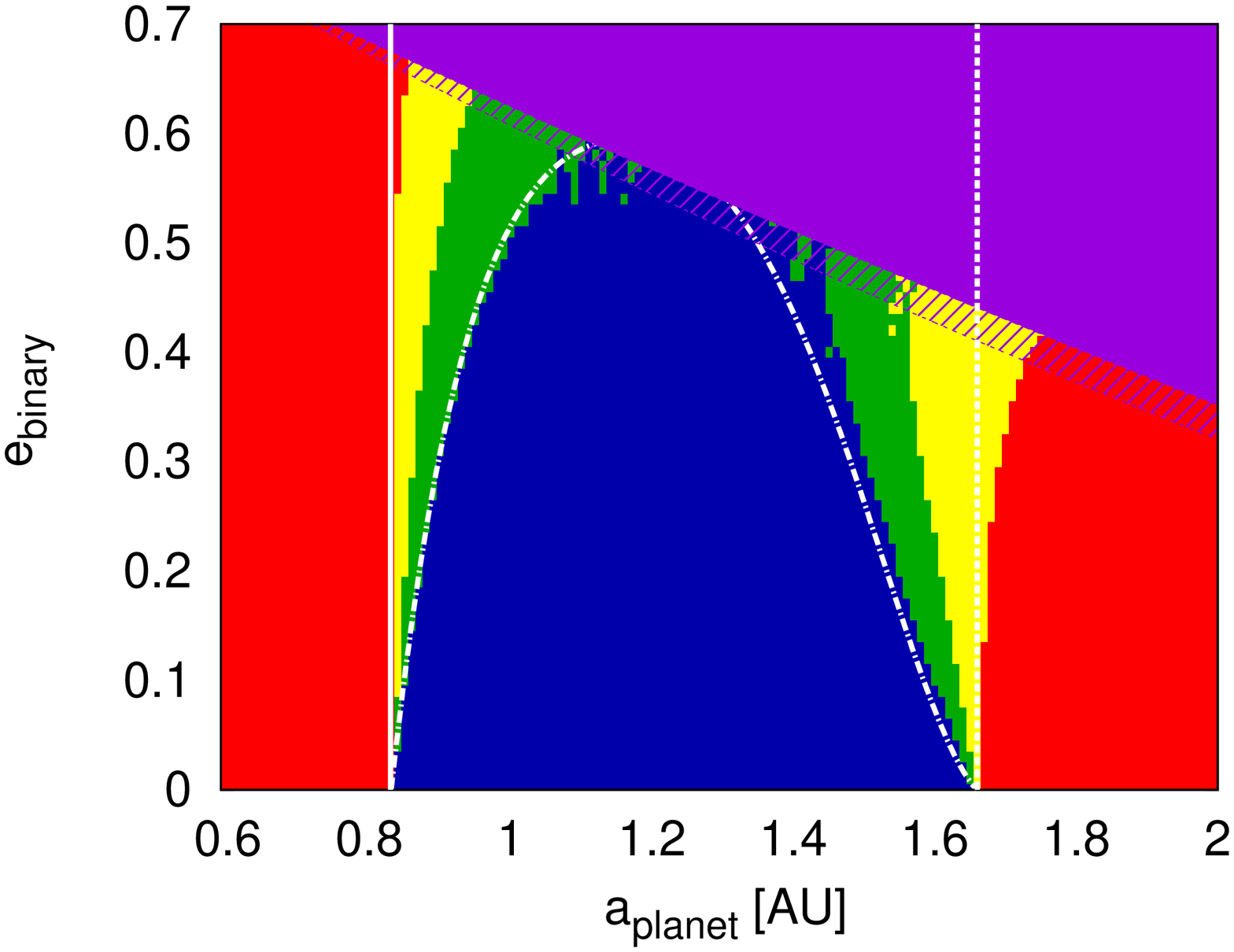}\\
\end{tabular}
\caption{Same as Fig.\,\ref{fig5} but for binary configurations with semi-major axis of $a_{b} = 10\;AU$.
A decrease in the binaries' semi-major axes leads to more pronounced differences between analytic estimates and numerical simulation results. 
This can be expected from the approximations used to calculate PHZ, EHZ and AHZ, see section~\ref{sec:ana}. 
Strong perturbations near the area of instability (gray striped - purple online) are modeled less accurately by the analytical estimates.
The white dashed-dotted line corresponds to the semi-analytic estimates of the PHZ using numerically derived values for $e_p^{max}$.  
 \label{fig6}}
\end{figure}

\end{document}